\documentclass[prb,twocolumn,showpacs]{revtex4}
\usepackage{graphicx}
\usepackage{amssymb,amsmath}

\begin{document}

\title{$H = x p$ with interaction and the Riemann zeros
}

\author{Germ\'an Sierra}

\affiliation{Instituto de F\'{\i}sica Te\'orica, CSIC-UAM, Madrid, Spain}
\date{February, 2007}

\bigskip\bigskip\bigskip\bigskip

%
\font\numbers=cmss12
\font\upright=cmu10 scaled\magstep1
\def\stroke{\vrule height8pt width0.4pt depth-0.1pt}
\def\topfleck{\vrule height8pt width0.5pt depth-5.9pt}
\def\botfleck{\vrule height2pt width0.5pt depth0.1pt}
\def\Zmath{\vcenter{\hbox{\numbers\rlap{\rlap{Z}\kern
0.8pt\topfleck}\kern 2.2pt
                   \rlap Z\kern 6pt\botfleck\kern 1pt}}}
\def\Qmath{\vcenter{\hbox{\upright\rlap{\rlap{Q}\kern
                   3.8pt\stroke}\phantom{Q}}}}
\def\Nmath{\vcenter{\hbox{\upright\rlap{I}\kern 1.7pt N}}}
\def\Cmath{\vcenter{\hbox{\upright\rlap{\rlap{C}\kern
                   3.8pt\stroke}\phantom{C}}}}
\def\Rmath{\vcenter{\hbox{\upright\rlap{I}\kern 1.7pt R}}}
\def\Z{\ifmmode\Zmath\else$\Zmath$\fi}
\def\Q{\ifmmode\Qmath\else$\Qmath$\fi}
\def\N{\ifmmode\Nmath\else$\Nmath$\fi}
\def\C{\ifmmode\Cmath\else$\Cmath$\fi}
\def\R{\ifmmode\Rmath\else$\Rmath$\fi}
\def\H{{\cal H}}
\def\NN{{\cal N}}

\begin{abstract}
Starting from a quantized version of the classical Hamiltonian 
$H = x p$, we add a non local interaction which depends
on two potentials. 
The model is solved exactly in terms of a Jost
like function which is analytic in the complex upper half plane. 
This function vanishes, either
on the real axis, corresponding to bound states, or below 
it, corresponding to resonances. 
We find potentials for which the 
resonances converge asymptotically toward
the average position of the Riemann zeros.  
These potentials realize, at the quantum level, the
semiclassical regularization of $H = x p$ 
proposed by Berry and Keating. Furthermore, a linear
superposition of them, obtained 
by the action of integer dilations, yields
a Jost function whose real part vanishes at the Riemann zeros
and whose imaginary part resembles the one of the zeta function. 
Our results suggest the existence of a quantum mechanical model
where the Riemann zeros would make a  point like spectrum embbeded
in the continuum. The associated spectral interpretation would
resolve the emission/absortion debate between Berry-Keating
and Connes. Finally, we indicate how our results
can be extended to the Dirichlet L-functions
constructed with real characters. 
\end{abstract}


\maketitle

\vskip 0.2cm

%
%
%
%
\def\oti{{\otimes}}
\def\lb{ \left[ }
\def\rb{ \right]  }
\def\tilde{\widetilde}
\def\bar{\overline}
\def\hat{\widehat}
\def\*{\star}
\def\[{\left[}
\def\]{\right]}
\def\({\left(}      \def\BL{\Bigr(}
\def\){\right)}     \def\BR{\Bigr)}
    \def\BBL{\lb}
    \def\BBR{\rb}
%
%
\def\zb{{\bar{z} }}
\def\zbar{{\bar{z} }}
\def\frac#1#2{{#1 \over #2}}
\def\inv#1{{1 \over #1}}
\def\half{{1 \over 2}}
\def\d{\partial}
\def\der#1{{\partial \over \partial #1}}
\def\dd#1#2{{\partial #1 \over \partial #2}}
\def\vev#1{\langle #1 \rangle}
\def\ket#1{ | #1 \rangle}
\def\rvac{\hbox{$\vert 0\rangle$}}
\def\lvac{\hbox{$\langle 0 \vert $}}
\def\2pi{\hbox{$2\pi i$}}
\def\e#1{{\rm e}^{^{\textstyle #1}}}
\def\grad#1{\,\nabla\!_{{#1}}\,}
\def\dsl{\raise.15ex\hbox{/}\kern-.57em\partial}
\def\Dsl{\,\raise.15ex\hbox{/}\mkern-.13.5mu D}
%
%
\def\ga{\gamma}     \def\Ga{\Gamma}
\def\be{\beta}
\def\al{\alpha}
\def\ep{\epsilon}
\def\vep{\varepsilon}
\def\dep{d}
\def\arc{{\rm Arctan}}
\def\la{\lambda}    \def\La{\Lambda}
\def\de{\delta}     \def\De{\Delta}
\def\om{\omega}     \def\Om{\Omega}
\def\sig{\sigma}    \def\Sig{\Sigma}
\def\vphi{\varphi}
\def\sign{{\rm sign}}
\def\he{\hat{e}}
\def\hf{\hat{f}}
\def\hg{\hat{g}}
\def\ha{\hat{a}}
\def\hb{\hat{b}}
%
%
\def\CA{{\cal A}}   \def\CB{{\cal B}}   \def\CC{{\cal C}}
\def\CD{{\cal D}}   \def\CE{{\cal E}}   \def\CF{{\cal F}}
\def\CG{{\cal G}}   \def\CH{{\cal H}}   \def\CI{{\cal J}}
\def\CJ{{\cal J}}   \def\CK{{\cal K}}   \def\CL{{\cal L}}
\def\CM{{\cal M}}   \def\CN{{\cal N}}   \def\CO{{\cal O}}
\def\CP{{\cal P}}   \def\CQ{{\cal Q}}   \def\CR{{\cal R}}
\def\CS{{\cal S}}   \def\CT{{\cal T}}   \def\CU{{\cal U}}
\def\CV{{\cal V}}   \def\CW{{\cal W}}   \def\CX{{\cal X}}
\def\CY{{\cal Y}}   \def\CZ{{\cal Z}}

\def\Hp{{\mathbb{H}^2_+}} 
\def\Hm{{\mathbb{H}^2_-}}

\def\rvac{\hbox{$\vert 0\rangle$}}
\def\lvac{\hbox{$\langle 0 \vert $}}
\def\comm#1#2{ \BBL\ #1\ ,\ #2 \BBR }
\def\2pi{\hbox{$2\pi i$}}
\def\e#1{{\rm e}^{^{\textstyle #1}}}
\def\grad#1{\,\nabla\!_{{#1}}\,}
\def\dsl{\raise.15ex\hbox{/}\kern-.57em\partial}
\def\Dsl{\,\raise.15ex\hbox{/}\mkern-.13.5mu D}
%
%
%
\font\numbers=cmss12
\font\upright=cmu10 scaled\magstep1
\def\stroke{\vrule height8pt width0.4pt depth-0.1pt}
\def\topfleck{\vrule height8pt width0.5pt depth-5.9pt}
\def\botfleck{\vrule height2pt width0.5pt depth0.1pt}
\def\Zmath{\vcenter{\hbox{\numbers\rlap{\rlap{Z}\kern
0.8pt\topfleck}\kern 2.2pt
                   \rlap Z\kern 6pt\botfleck\kern 1pt}}}
\def\Qmath{\vcenter{\hbox{\upright\rlap{\rlap{Q}\kern
                   3.8pt\stroke}\phantom{Q}}}}
\def\Nmath{\vcenter{\hbox{\upright\rlap{I}\kern 1.7pt N}}}
\def\Cmath{\vcenter{\hbox{\upright\rlap{\rlap{C}\kern
                   3.8pt\stroke}\phantom{C}}}}
\def\Rmath{\vcenter{\hbox{\upright\rlap{I}\kern 1.7pt R}}}
\def\Z{\ifmmode\Zmath\else$\Zmath$\fi}
\def\Q{\ifmmode\Qmath\else$\Qmath$\fi}
\def\N{\ifmmode\Nmath\else$\Nmath$\fi}
\def\C{\ifmmode\Cmath\else$\Cmath$\fi}
\def\R{\ifmmode\Rmath\else$\Rmath$\fi}

\def\barray{\begin{eqnarray}}
\def\earray{\end{eqnarray}}
\def\beq{\begin{equation}}
\def\eeq{\end{equation}}

\def\no{\noindent}

\def\gpar{g_\parallel}
\def\gperp{g_\perp}

\def\Jb{\bar{J}}
\def\dx{\frac{d^2 x}{2\pi}}

\def\rap{\beta}
\def\s{\sigma}
\def\spec{\zeta}
\def\comb{\frac{\rap\theta}{2\pi} }
\def\Ga{\Gamma}

\def\L{{\cal L}}
\def\g{{\bf g}}
\def\K{{\cal K}}
\def\I{{\cal I}}
\def\M{{\cal M}}
\def\F{{\cal F}}

\def\gpar{g_\parallel}
\def\gperp{g_\perp}
\def\Jb{\bar{J}}
\def\dx{\frac{d^2 x}{2\pi}}
\def\imag{\Im {\it m}}
\def\real{\Re {\it e}}
\def\Jbar{{\bar{J}}}
\def\kh{{\hat{k}}}
\def\Im{{\rm Im}}
\def\Re{{\rm Re}}
\def\ti{{\tilde{\phi}}}
\def\tR{{\tilde{R}}}
\def\tS{{\tilde{S}}}
\def\tF{{\tilde{\cal F}}}

\section{Introduction}

The Riemann hypothesis is considered 
the most important problem in Analytic
Number Theory \cite{Edwards,Titchmarsh2,Bombieri,Sarnak,Conrey}. 
It states that the non trivial zeros of the
classical zeta function have real part equal to $1/2$. 
Hilbert and P\'olya suggested long ago
that the RH can be proved if one finds
a self-adjoint linear operator whose 
eigenvalues are the Riemann zeros
\cite{Watkins,Rosu,Elizalde}. 
The first indication of the adecuacy
of this conjecture was probably the work by Selberg in the 1950s,
who found a remarkable duality between the 
eigenvalues of the Laplacian acting on Riemann
surfaces of constant negative curvature 
and the length spectrum of their geodesics \cite{Selberg}. 
Selberg trace formula, which establishes that link, 
strongly resembles Riemann explicit formula. 
Another important hint 
came in 1973 from  Montgomery's  work who, 
assuming the RH, 
 showed that the Riemann zeros are distributed according to the  
Gaussian Unitary Ensemble statistics
of random matrix models \cite{Mont}. Montgomery's  results, 
were  confirmed by the impressive numerical
findings obtained by Odlyzko in the 1980's \cite{Odl}.  
The next step in this direction
was put forward by Berry 
who proposed the Quantum Chaos conjecture,
according to which the Riemann zeros
are the spectrum of a Hamiltonian
obtained by quantization of a classical 
chaotic Hamiltonian, whose periodic
orbits are labelled by the prime numbers \cite{B-chaos}. 
This suggestion
was based on analogies between fluctuation formulae
in Number Theory and  Quantum Chaos \cite{Gutzwiller}. 
Another interesting approaches to the RH are
based on Statistical Mechanical ideas
\cite{Julia,BC}. The prime numbers has also been considered
from a quantum mechanical viewpoint \cite{Mussardo}.

Up to date, it is not known a Hamiltonian  
accomplishing the Hilbert-P\'olya conjecture. 
Along these lines, Berry and Keating 
suggested in 1999 that the 1d classical 
Hamiltonian $H = x p$ is related to the Riemann
zeros \cite{BK1,BK2}. This suggestion was based on a heuristic and 
semiclassical analysis which yields, rather surprisingly, the average
number of  Riemann zeros up to a given height. 
Unfortunately, this encouraging 
result does not have a quantum counterpart. More explicitely,
it is not known a quantization of $H = x p$ 
yielding the average, or exact, position of the Riemann zeros
as eigenvalues. The Berry-Keating papers
were  inspired by an earlier one from Connes who 
tried to prove  the RH in terms of the mathematical structures
known as adeles and p-adic numbers \cite{Connes}. In order to illustrate
the adelic approach, Connes introduced the Hamiltonian
$H = xp$,  using a different semiclassical regularization.
In Connes's approach the  Riemann zeros appear
as missing spectral lines in a continuum, which
does not  conform to  the Berry-Keating's  approach
where the Riemann zeros appear as discrete spectra.
Both approaches are heuristic and semiclassical, 
therefore the apparent contradiction between them  
cannot be resolved until one derives a consistent
quantum theory of $H = x p$, and its possible
extensions. 

In reference \cite{JSTAT} we proposed a quantization of 
$H = x p$ using an unexpected connection of this
model to the one-body version of the so called Russian
doll  BCS model of superconductivity \cite{RD1,RD2,links}. 
The latter model was, in turn, motivated 
by previous papers on the Renormalization 
Group with limit cycles \cite{GW,BLflow,nuclear} 
(see also \cite{fewbody,morozov}). 
The relation between $H = x p$ and the Russian doll (RD) model 
is as follows. An eigenstate, with energy $E$, of a quantum version
of the classical Hamiltonian $H = x p$, corresponds to a zero
energy eigenstate of the RD Hamiltonian, where $E$ becomes
a  coupling constant. Since the RD model is exactly
solvable \cite{links}, so it is the $H = x p$ model. The spectrum
obtained in this way was shown to agree 
with Connes's picture of a continuum of eigenstates \cite{JSTAT}. 
We also obtained the smooth part of the Riemann
formula for the zeros, however this fact cannot be interpreted
as missing states but rather as a blueshift
of energy levels.  A point like spectrum associated to the Riemann
zeros was completely absent in this quantization 
of $H = x p$.  The final conclusion of 
\cite{JSTAT} was the necessity to go beyond the $H = x p$ model, 
in order to realize an spectral interpretation of the Riemann
zeros. Some proposals were already made in that reference  
but the corresponding models could not be solved 
exactly. 

The cyclic Renormalization Group, and its realization
in the field theory models of references \cite{LRS1,LRS2,LS},
is at the origin of LeClair's approach to the RH \cite{Andre}.
In this reference the zeta function on the critical strip
is related to the quantum statistical mechanics of non-relativistic,
interacting fermionic gases in 1d with a quasi-periodic 
two-body potential. This quasi-periodicity is reminiscent
of the zero temperature 
cyclic RG of the quantum mechanical Hamiltonian of \cite{JSTAT}, 
but the general framework of both works is different.
The cyclic RG underlies several 
of the results of the present paper, but we shall
not deal with it in the rest of the paper.

The organization of the paper is as follows. 
In section II we  review
the Berry-Keating and Connes 
semiclassical approaches to $H = x p$. 
In section III we quantize this Hamiltonian,  
finding its self-adjoint extensions
and their relation to the semiclassical
approaches of section II. We also
study the inverse Hamiltonian
$1/( xp)$ and its connection to  the 
Russian doll model. 
In section IV we add an interaction
to a quantized version of 
$H = 1/( x p)$, and solve the general model exactly,
in terms of a Jost like function.
Section V is devoted to the analiticity
properties of this Jost function.  
In section VI we study the potentials
which exhibit some relation to the Riemann zeros.

\section{Semiclassical approach}

The classical Berry-Keating-Connes (BKC) Hamiltonian  
\cite{BK1,BK2,Connes}

\beq
H^{\rm cl}_{0} = x \;  p,  
\label{s1}
\eeq

\no 
has classical trayectories  given by the  hyperbolas 
(see fig.1a)

\beq
x(t) = x_0 \; e^{t} , \quad p(t) = p_0 \;  e^{-t}.  
\label{s2}
\eeq

\no
The dynamics is unbounded, so 
one should not expect a discrete spectrum at the
quantum level. In 1999  
Berry and Keating on the one hand \cite{BK1,BK2}, 
and Connes on the other \cite{Connes},
introduced two different types of regularizations
of the model and made a semiclassical 
counting of states. Berry and Keating proposed
the Planck cell in phase space:  $|x| > l_x$ and $|p| > l_p$, 
with $l_x \, l_p = 2 \pi \hbar$, while Connes choosed 
$|x| < \Lambda$ and $|p| < \Lambda$, where $\Lambda$ is a cutoff. 
In reference \cite{JSTAT} we considered a third 
regularization which combines the previous ones 
involving the position $x$, namely  
$l_x < x < \Lambda$, making  no assumption for the momenta $p$.
The number, $\CN(E)$, of semiclassical states
with an energy lying between $0$ and $E$ is given by

\beq
\CN(E) = \frac{A}{2 \pi \hbar}, 
\label{s3}
\eeq

\no
where $A$ is the area of the allowed phase space region 
below the curve $E = x \;p$ (see figs.1b,1c,1d). Table 1 
collects the values of $\CN(E)$ for the three types
of regularizations.

\begin{center}
\begin{tabular}{|c|c|c|}
\hline
Type & Regularization &   $\CN(E)$ \\
\hline 
\hline
BK& $ |x| > l_x, \; \;  |p| > l_p$ &  $\frac{E}{2 \pi} 
\left( \log \frac{E}{2 \pi} -1 \right) + 1$   \\
\hline
C &  $ |x| < \Lambda, \; \;  |p| < \Lambda$ & 
$ \frac{E}{ \pi} \log \Lambda
- \frac{E}{2 \pi} \left( \log \frac{E}{2 \pi} -1 \right) $ \\
\hline
S & $ l_x < x < \Lambda$ &   
$ \frac{E}{2 \pi} \log \frac{\Lambda}{l_x}$ \\
\hline
\end{tabular}

\vspace{0.5 cm}

Table 1.- Three different  
regularizations of $H = x p$ and the corresponding 
number of semiclassical states in units $\hbar =1$. 
\end{center}

\no 
In the BK regularization,  
the number of semiclassical states 
agrees,  quite remarkably, with the 
asymptotic limit of the smooth part of the formula
giving the 
number of  Riemann zeros whose imaginary part lies
in the interval $(0,E)$,

\beq
\langle \CN(E) \rangle \sim  \frac{E}{2 \pi} 
\left( \log \frac{E}{2 \pi} -1 \right) + \frac{7}{8} + \dots,\qquad E >> 1.
\label{s4} 
\eeq

\no
The exact formula for the number of zeros, $\CN_R(E)$, is due to Riemann,
and contains also a fluctuation term which depends on 
the zeta function \cite{Edwards},

\barray
\CN_R(E) & = & \langle \CN(E) \rangle  + N_{\rm fl}(E)  
\label{s5}
\\
\langle \CN(E) \rangle &  = & 
 \frac{1}{\pi} \, 
  \Im \log \Gamma \left( \frac{1}{4} + 
\frac{i}{2} E \right) 
- \frac{E}{2 \pi} \log \pi + 1 
\nonumber 
 \\
\CN_{\rm fl}(E) & = &  = \frac{1}{\pi} \Im \log \zeta \left( 
\frac{1}{2} + i E \right) \nonumber 
\earray

\begin{figure}[t!]
\begin{center}
\includegraphics[height= 8 cm,angle= 0]{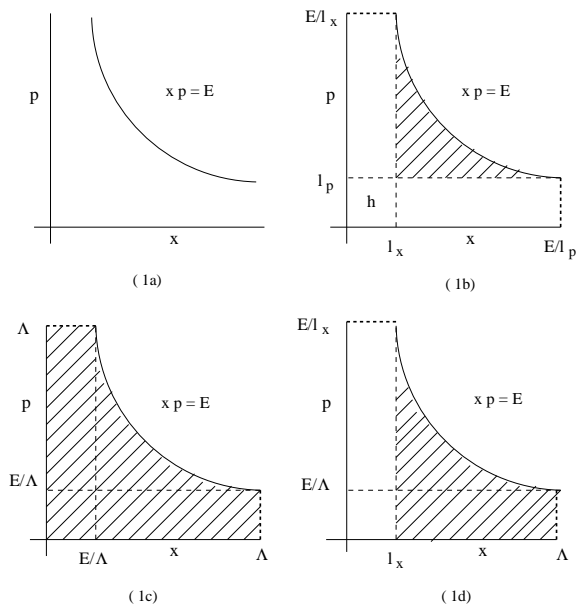}
\end{center}
\caption{
1a) a classical trayectory (\ref{s2}).  
The regions in shadow are the allowed 
phase space of the semiclassical regularizations
of $H = x p$ considered 
by:  1b) Berry and Keating, 1c) Connes and 1d) Sierra. 
The values of the associated areas are given in table 1.  
}
\label{semiclassical}
\end{figure}

\no
Based on this result, and  analogies
between formulae in Number Theory and Quantum Chaos,  
Berry and Keating suggested the existence
of a classical chaotic hamiltonian 
whose quantization would give rise  to the zeros
as point like spectra \cite{BK1,BK2}. 
They conjectured the properties of this classical
Hamiltonian, 
which include the breaking of time reversal
symmetry, which holds for (\ref{s1}),
and the existence of primitive periodic
orbits labelled by the prime numbers. 
However, up to now, there is no a concrete proposal
realizing all these conditions. 

On the other hand,  
Connes  found 
that the number of semiclassical states diverges 
in the limit
where the cutoff $\Lambda$ goes to infinity, and that
there is a finite size correction given by
{\em minus} the average position of the Riemann zeros
(see table 1). This result led 
to the  missing spectral interpretation of the 
Riemann zeros, according to which there is a continuum
of eigenstates (represented by the term 
$\frac{E}{ \pi} \log \Lambda$ in $\CN(E)$), 
where some of the states are missing, precisely the ones associated
to the Riemann zeros. This interpretation, albeit  appealing,
has the trouble 
that the number of missing states changes  
linearly in $E$ after scaling the cutoff $\Lambda$,
and thus it is regularization dependent.  
As in the BK case, 
the C-regularization is not supported by a quantum mechanical
version of $H = x p$, although it serves to illustrate, in a simple
example, the main ideas underlying the Connes's adelic approach
to the Riemann hypothesis.

Finally, in the S-regularization the number of semiclassical
states diverges as  $\frac{E}{ 2 \pi} \log \Lambda/l_x$, suggesting
a continuum spectrum, like  in Connes's approach. But there
is no a finite size correction to that formula, and consequently
the possible connection to the Riemann zeros is lost. The main
advantage of this regularization is that the Hamiltonian
(\ref{s1}) can be consistently quantized  yielding
a  spectrum which coincides with the semiclassical result
as we show below.

\section{Quantization of $ x \, p$ and $1/( x p)$}

\subsection{The  Hamiltonian $H_0 = x p$}

In this section we construct 
a self-adjoint operator $H_0$, associated to 
$H_0^{\rm cl} = x \, p$, which acts on the Hilbert
space  $L^2(a,b)$ of square integrable
functions  in the interval $(a,b)$. Assuming that
$x  \geq 0$, there are four possible intervals
corresponding to the choices:
$a = 0, l_x$ and $b = \Lambda, \infty$, where
$l_x$ and $\Lambda$ were introduced above (we shall take 
 $l_x = 1$ and $\Lambda = N > 1$). 
Berry and Keating defined the  quantum Hamiltonian $H_0$ 
as the normal ordered expression

\beq
H_0 = \frac{1}{2} ( x \, p + p \, x),
\label{qu1}
\eeq

\no
where $ p = - i \hbar \;  d/dx$. If $x \geq 0$, eq.(\ref{qu1})
is equivalent to

\beq
H_0 = \sqrt{x} \, p \,  \sqrt{x} = - i \hbar \sqrt{x} \, \frac{d}{d x} 
\,  \sqrt{x}. 
\label{qu2}
\eeq

\no 
This is a symmetric operator acting in a certain domain of the
Hilbert space  $L^2(a,b)$, if \cite{Galindo}

\beq
\langle \psi| H_0 \phi \rangle - \langle H_0 \psi|  \phi \rangle 
= i \hbar \[ a \psi^*(a) \phi(a) - b \psi^*(b) \phi(b) \] = 0,  
\label{qu3}
\eeq

\no
which is satisfied if both $\psi(x)$ and $\phi(x)$ 
vanish at the points $a, b$.  By a theorem due
to von Neumman, the symmetric operator
$H_0$ is also self-adjoint if 
its deficiency indices $n_\pm$ are equal  \cite{Neumann}.
These indices counts the number of solutions of the eq.

\beq
H_0^\dagger \,  \psi_\pm  = \pm i \hbar \, \lambda \, \psi_\pm, 
\label{qu4}
\eeq

\no
belonging to the domain of $H_0^\dagger$ ($\lambda > 0$). 
If $n= n_+ = n_- > 0$, 
there are infinitely many self-adjoint extensions of $H_0$ parameterized
by a unitary $n \times n$ matrix. The solutions of eq.(\ref{qu4})
are 

\beq
\psi_\pm(x)  = C  x^{- 1/2  \mp \lambda},  
\label{qu5}
\eeq

\no
whose norm in the interval $(a,b)$ is, 

\beq
\langle \psi_\pm | \psi_\pm \rangle = 
\pm \frac{C^2}{2 \lambda} (a^{ \mp 2 \lambda} - b^{ \mp 2 \lambda}). 
\label{qu6}
\eeq

\no 
The deficiency indices corresponding to the four intervals considered
above are collected in table 2.

\begin{center}
\begin{tabular}{|c|c|c|c|}
\hline
Type & $(a,b)$ & $(n_+, n_-)$ & Self-adjoint \\
\hline 
BK & $(1,\infty)$ & $(1,0)$ & -   \\
\hline 
C & $(0, N)$ & $(0,1)$ &  - \\
\hline
S & $(1, N)$ & $(1,1)$ & $\surd$  \\
\hline 
T & $(0, \infty)$ & $(0,0)$ &  $\surd$ \\
\hline
\end{tabular}

\vspace{0.5 cm}

Table 2.- Deficiency indices of $H_0$. The
corresponding intervals are associated
to the semiclassical regularizations
of section II (i.e. BK, C, S). The last one, 
T, describes the case with no constraints
on $x$ except positivity (i.e. $x > 0)$.
\end{center}

The von Neumann
theorem implies that the operator $H_0$ 
is essentially self-adjoint
on the half line $\R_+ = (0,\infty)$. 
This case was recently studied by Twamley and Milburn,
who defined a  quantum Mellin transform using the
eigenstates of $H_0$  \cite{Twamley}.

On the other hand, in the interval  $(1,N)$ the operator $H_0$ 
admits  infinitely many self-adjoint
extensions parameterized by a phase $e^{i \theta}$. 
This phase determines the boundary conditions of the functions
belonging to the self-adjoint domain

\beq
{\cal D} (H_{0, \theta}) = \left\{ \psi, H_0 \psi  \in  L^2(1,N), \;\; 
  e^{i \theta} \psi(1)  = \sqrt{N} \psi(N) \right\}. 
\label{qu7}
\eeq

\no
The eigenfunctions of $H_0$,

\beq
H_0 \; \psi_E = E \; \psi_E ,
\label{qu8}
\eeq

\no
are given by \cite{BK1}

\beq
\psi_E(x) = \frac{C}{x^{1/2 - i E \hbar}}, \qquad E \in \Rmath,  
\label{qu9}
\eeq

\no
where $C$ is a normalization constant. In the half line
$\R_+$ there are no further restrictions on $E$, hence the
spectrum  of $H_0$ is continuous 
and covers the whole real line $\Rmath$. 
In this case the normalization constant in (\ref{qu9})
is choosen as $C = 1/\sqrt{2 \pi \hbar}$ which
guarantees the standard normalization

\beq
\langle \psi_E| \psi_{E'} \rangle
= C^2 \int_0^\infty \frac{dx}{x} x^{- i (E - E')/\hbar} = \delta(E - E').
\label{qu10}
\eeq

\no
In the case where $H_0$ is defined in the interval $(1,N)$,
the boundary condition (\ref{qu7}) yields
the quantization condition for $E$, namely

\beq
N^{i E/\hbar} = e^{i \theta} \Longrightarrow
E_n = \frac{2 \pi \hbar}{\log N} \left( n + \frac{\theta}{2 \pi}
\right), \;\; n \in \N. 
\label{qu11}
\eeq

\no
Hence the spectrum of $H_0$ is discrete, with a
level spacing decreasing  for large values of $N$. 
The normalization constant of the wave function
is  now $C = \frac{1}{\sqrt{\log N}}$
which gives,

\beq
\langle \psi_{E_n}| \psi_{E_{n'}} \rangle
= C^2 \int_1^N \frac{dx}{x} x^{- i (E_n - E_{n'})/\hbar} = \delta_{n,n'}.
\label{qu12}
\eeq

\no
The spectrum (\ref{qu11})  agrees 
with the semiclassical result given in table 1 for the
S-regularization (recall that $l_x =1, \Lambda =  N, \hbar =1)$.

The existence of only two self-adjoint
extensions of the operator $H_0$, in the positive
real axis, should not be surprising, since
they are intimately related to those of 
the momenta operator $P = - i \hbar \frac{d}{dq}$, 
where $q = \log x$. Indeed,  the $P$ operator
defined on $\Rmath$, admits only two
self-adjoint extensions in the $q$-intervals: 
$(-\infty,\infty)$ and $(a,b)$ ($a, b$ finite),
which correspond to the $x$-intervals:
$(0,\infty)$ and $(\log a, \log b)$, respectively. 
Under this mapping, the wave function
(\ref{qu9}) corresponds to the plane wave
$e^{i q E}$, where $x^{1/2}$
is a measure factor. The spectrum of $H_0$ 
can therefore  be understood
in terms of the familiar spectrum of $P$.

Returning to (\ref{qu11}), for the particular case where $\theta = \pi$, 
one observes that the energy spectrum
is symmetric around zero, i.e. if $E_n$ is an eigenenergy
so is $- E_n$. This result was obtained in reference
\cite{JSTAT} working with the inverse Hamiltonian
$1/H_0$. We shall next review that construction since
it will be important in the sequel.

\subsection{The inverse Hamiltonian $1/H_0$}

First, we start from the expression (\ref{qu2}) 
and take the formal 
inverse, i.e. $ H_0^{-1} = x^{-1/2} \, p^{-1} \,  x^{-1/2}$. 
The operator $p^{-1}$ is the one dimensional Green function
with matrix elements $\langle x| p^{-1} |x' \rangle =
 \frac{i}{2 \hbar} \;  \sign(x-x')$, where  
$\sign(x-x')$ is the sign function. 
The operator $H_0^{-1}$ is defined in the interval
$(1,N)$  by the continuous matrix,

\beq
H_0^{-1}(x,x') = \frac{i}{2 \hbar} \frac{\sign(x - x')}{\sqrt{x \, x'}},
\qquad 1 \leq x, x' \leq N.  
\label{qu13}
\eeq

\no
Its spectrum is found solving the Schr\"odinger equation

\beq
\frac{i}{2 \hbar}
\int_1^N d x' \;  \frac{\sign(x - x')}{\sqrt{x \, x'}} \;
\psi(x') = E^{-1} \;  \psi(x),  
\label{qu14}
\eeq

\no
for the eigenvalue $E^{-1}$, which must not be singular
for  $H_0^{-1}$ to be invertible. Define
a new wave function 

\beq
\phi(x) = \frac{\psi(x)}{\sqrt{x}},  
\label{qu15}
\eeq

\no
which satisfies

\beq
\frac{i E}{2 \hbar}
\int_1^N d x' \;  \sign(x - x') \; 
\phi(x') =  x \;  \phi(x).  
\label{qu16}
\eeq

\no
Taking the derivative with respect to $x$ yields

\beq
x \frac{d }{d x} \phi(x) = \left( 1 - \frac{i E}{\hbar} \right)
\phi(x),    
\label{qu17}
\eeq

\no
which is solved by

\beq
\phi(x) = \frac{C}{x^{1 - i E/\hbar}}  
\Rightarrow \psi(x) = \frac{C}{x^{1/2  -  i  E/\hbar}},   
\label{qu18}
\eeq

\no
with $C = 1/\sqrt{\log N}$ as in (\ref{qu12}). 
Eq. (\ref{qu18}) fixes
the functional form of $\psi(x)$.
To find the spectrum we impose 
(\ref{qu16}) at one point, say $x =1$, 
obtaining,

\beq
N^{i E/\hbar} = -1 \Longrightarrow 
E_n = \frac{2 \pi \hbar}{\log N} ( n + \frac{1}{2} ), \qquad
n \in \Nmath 
\label{qu20}
\eeq

\no
This spectrum coincides with (\ref{qu11})
for $\theta = \pi$, so that
the eigenenergies come in pairs
 $\{ E_n, -E_n \}$,  as corresponds to an hermitean
antisymmetric operator. Including a BCS coupling 
in (\ref{qu13}), related to $\theta$, 
yields the spectrum (\ref{qu11}) \cite{JSTAT}.

In summary, we have constructed in this section
a quantum version of the classical Hamiltonian
$x p$, as well as its inverse $1/( x p)$, which
agree with the semiclassical regularization
$ l_x < x < \Lambda$.  
Both, the semiclassical regularization,  
and the associated quantization, 
shows no trace of the Riemann zeros, 
which suggests that a possible connection
to them requires to go beyond the $x p$ model. In the
next section we shall take a further step in that direction.

\section{$H_0 = x \;p$  with interactions}

\subsection{Definition of the Hamiltonian}

The standard way to add an interaction to a free Hamiltonian
$H_0$ is to perturb it by a potential term, i.e.

\beq
H = H_0 + V.
\label{c0} 
\eeq

\no
Instead of starting from  the Hamiltonian $H_0$ (\ref{qu1}) 
we shall perturb the inverse
Hamiltonian $1/H_0$ (\ref{qu13})

\beq
\frac{1}{H} = \frac{1}{H_0} + V'
\label{c01} 
\eeq

\no 
so that $H$ depends  non linearly on $V'$.
We have found more convenient to work with
(\ref{c01}), rather than with (\ref{c0}) but,  
of course, the two formulations must be related
(we leave this issue for a later publication). 

The interacting Hamiltonian $1/H$ that
we shall consider is given by,  

\beq
H^{-1}_2(x,x') = \frac{i}{2 \hbar} 
\frac{ \sign(x-x') + a(x) b(x') - b(x) a(x') }{ \sqrt{\vep(x) \vep(x')}}, 
\label{c1} 
\eeq

\no
where $a(x)$ and $b(x)$ are two real functions defined in the
interval $x \in (1, N)$, and $\vep(x)$ is a positive and 
monotonically increasing function. The BKC model
corresponds to the choice $\vep(x) = x$, but it is 
equally easy 
to work with generic functions  $\vep(x)$, which 
also links the present
model to the RD model,  where 
$\vep(x)$ gives the energy levels of electrons pairs.

$H^{-1}_2$  is an hermitean antisymmetric operator, and hence its spectrum
is real and symmetric around zero. We shall assume, for the time being, that
$H^{-1}_2$ is invertible, a condition which depends on the potentials
 $a(x)$ and $b(x)$. If one of the potentials is constant, say 
$b(x) = 1$, then  (\ref{c1}) becomes

\beq
H^{-1}_1(x,x') = \frac{i}{2 \hbar} 
\frac{ \sign(x-x') + a(x)  -  a(x') }{ \sqrt{\vep(x) \vep(x')}}.
\label{c2} 
\eeq

\no
We shall denote $\M_2$ (resp. $\M_1$) 
the model with Hamiltonian (\ref{c1}) (reps. (\ref{c2})). 
These two  models share many properties but they differ in some
important instances. For example, (\ref{c1}) is invariant under the  
transformation

\beq
\left(
\begin{array}{cc}
a(x) \\
b(x) \\
\end{array}
\right)  \rightarrow 
\left(
\begin{array}{cc}
\alpha & \beta  \\
\gamma & \delta \\
\end{array}
\right) \left(
\begin{array}{cc}
a(x) \\
b(x) \\
\end{array}
\right), \qquad \forall x \in (1,N),  
\label{c3} 
\eeq

\no where the $2 \times 2$ matrix is an element
of the $Sl(2, \R)$ group,

\beq
\left(
\begin{array}{cc}
\alpha & \beta  \\
\gamma & \delta \\
\end{array}
\right) \in Sl(2, \R) \Leftrightarrow \alpha, \beta, \gamma, \delta \in 
\R, \;\;
\alpha \delta - \beta \gamma = 1. 
\label{c4} 
\eeq

\no
while (\ref{c2}) is invariant under the
translations,

\beq
a(x) \rightarrow a(x) + \alpha,  \qquad \alpha \in \R. 
\label{c5} 
\eeq

\no
Naturally, the spectra of  $H_{1,2}$
must be invariant under the corresponding symmetry transformations. 
The Hamiltonian (\ref{c1}) can also be written as

\beq
H_2^{-1}(x,x') = H_0^{-1}(x,x') + \frac{i}{2 \hbar}
\left[ \psi_a(x) \psi_b(x') -  \psi_b(x) \psi_a(x') \right], 
\label{c1-1}
\eeq

\no
where 

\beq
\psi_a(x) = \frac{a(x)}{\sqrt{\vep(x)}}, \quad
\psi_b(x) = \frac{b(x)}{\sqrt{\vep(x)}}. 
\label{c1-4}
\eeq

\no
This means that the interaction is given
by a sort of proyection operator
formed by the states  $\psi_{a,b}$. 
In section VI we show that
a particular choice of  $\psi_{a,b}$
provides a quantum version of the BK semiclassical
regularization conditions. The Hamiltonians
(\ref{c1}) and (\ref{c2}) admit a discrete
version analogue to the one considered
in \cite{JSTAT}. The results we shall derive 
in the coming sections are also valid in this
case. Furthermore, the connection with the RD model provides
an interesting many-body generalization which will
studied in a separate work.

\subsection{Solution of the Schr\"odinger equation}

The Schr\"odinger equation associated to (\ref{c1})
reads (in units of $\hbar = 1$),

\beq
\frac{i}{2} \int_1^N dx' \; 
\frac{ \sign(x-x') + a(x) b(x') - b(x) a(x') }{ \sqrt{\vep(x) \vep(x')}}
=  E^{-1}  \psi(x) 
 \label{c6} 
\eeq

\no
which for the wave function

\beq
\phi(x) = \frac{ \psi(x)}{\sqrt{ \vep(x)}}, 
\label{c7} 
\eeq

\no
becomes

\barray
& \vep(x)  \; \phi(x) =   \label{c8} & \\
& \frac{i E}{2} \int_1^N dx' \; 
\left(  \sign(x-x') + a(x) b(x') - b(x) a(x') \right) 
\phi(x'). & \nonumber   
\earray

\no
This equation is the basis for the relation between
the BKC and the RD models. Indeed, defining the RD Hamiltonian

\barray
& H_{RD_2}(x,x')   = &  \label{c9}  \\
& \vep(x) \delta(x-x') - 
\frac{i h_D}{2} 
\left(  \sign(x-x') + a(x) b(x') - b(x) a(x') \right),  
& \nonumber 
\earray

\no
we see that (\ref{c8}) becomes the eigenequation  
of a zero energy eigenstate $\phi(x)$,

\beq
H_2 \; | \psi \rangle = E  | \psi \rangle 
\Longleftrightarrow 
H_{RD_2}  \;  | \phi \rangle = 0, 
\label{c9b} 
\eeq

\no
provided the coupling
$h_D$ is related to the energy $E$ by

\beq
h_D = E.   
\label{c10} 
\eeq

\no
Eqs.(\ref{c9b}) and (\ref{c10}) establish a  one-to-one correspondence
between the energy spectrum of the Hamiltonian
$H_2$,  and the coupling constant spectrum  of  zero
energy states of the Hamiltonian
$H_{RD_2}$. In this regard, we shall  mention 
the work by Khuri \cite{khuri}, 
based on a suggestion by Chadan \cite{Chadan},  
where the Riemann zeros are  related to the ``coupling constant
spectrum'' of zero energy, s-wave, scattering problem for repulsive
potentials in standard Quantum Mechanics. In that model the
coupling constant, $\lambda$, is related to the zeros,
$s_n = 1/2 + i \gamma_n$, by the quadratic equation $\lambda = s ( s-1)$. 
In our model however, the relation between the coupling constant, $h_{RD}$,
and the energy $E$ is linear (eq.(\ref{c10})), which
is due to the fact that $H_0$ depends linearly on $d/dx$, 
while in standard QM  the kinetic term depends quadratically.

We shall next solve eq.(\ref{c8}) for generic values of $a(x)$
and $b(x)$. Later on, 
we shall impose additional constraints on these functions
so that the model makes sense in the limit $N \rightarrow \infty$. 
First of all, let us write (\ref{c8}) as

\beq
 \frac{i E}{2} \int_1^N dx' \; 
\sign(x-x') \; \phi(x') + a(x) B - b(x) A  = \vep(x)  \; \phi(x),  
\label{c11} 
\eeq

\no
where

\beq
A =  \frac{i E}{2} \int_1^N dx \; 
a(x) \;  \phi(x), \; 
B =  \frac{i E}{2} \int_1^N dx \; 
b(x) \;  \phi(x). \qquad 
\label{c12} 
\eeq

\no
Eq.(\ref{c11}) is equivalent to

\barray 
& i E  \phi(x) +  \frac{d a}{d x}  B - \frac{d b}{d x} A   
 =  \frac{d}{dx} \left( \vep(x)  \; \phi(x) \right) & 
\label{c13} \\
& -  \frac{i E}{2} \int_1^N dx \; \phi(x) + 
a_1 B - b_1 A    =   \vep(1)  \; \phi(1), &  
\label{c13b} 
\earray

\no 
which are obtained from (\ref{c11})
by taking the derivative  with respect to $x$, 
and setting $x=1$ with $a_1 = a(x=1)$ and $b_1 = b(x=1)$. 
Define the  variable $q$

\beq
q = \int_1^x \frac{dx'}{ \vep(x')},  
\label{c14} 
\eeq

\no
such that

\beq
q \in (0, L_N), \qquad L_N =  \int_1^N \frac{dx'}{ \vep(x')}.  
\label{c15} 
\eeq

\no
In the BKC model, i.e. $\vep(x) = x$, one gets $q = \log x$
and $L_N = \log N$. For more general choices of $\vep(x)$
we shall assume that $L_N \rightarrow \infty$ when
$N \rightarrow \infty$. In terms of the new function

\beq
\ti(x) = \vep(x) \; \phi(x),  
\label{c16} 
\eeq

\no
eq.(\ref{c13}) turns into

\beq
\left( \frac{d}{d q} - i E \right) 
\ti(q) =  \frac{d a}{d q}  B - \frac{d b}{d q} A,   
\label{c17} 
\eeq

\no
where $a, b $ and $\ti$ are regarded as functions of $q$. 
If $a$ and $b$ are zero, the solution of (\ref{c17})
is the plane wave $C e^{i E q}$, with $C$ a constant.
Otherwise, $C$  depends on $q$ and satisfies

\beq
 \frac{d C}{d q} = e^{- i E q} \left( 
  \frac{d a}{d q}  B - \frac{d b}{d q} A  \right),  
\label{c18} 
\eeq

\no
whose solution is

\beq
C(q)  =  C + \int_0^q dq' \,   e^{- i E q'} \left( 
  \frac{d a}{d q'}  B - \frac{d b}{d q'} A  \right),  
\label{c19} 
\eeq

\no
where $C$ is an integration constant. This equation fixes
the functional form of $\ti(q)$, namely

\beq
\ti(q)   =  C e^{i E q} + e^{i E q}  \int_0^q dq'  e^{- i E q'} \left( 
  \frac{d a}{d q'}  B - \frac{d b}{d q'} A  \right),  
\label{c20} 
\eeq

\no
and in turn of $\psi(x)$
by means of (\ref{c7}) and (\ref{c16}),

\beq
\psi(q)   =   \frac{e^{i q E}}{\sqrt{\vep(q)}}
\left[   C  +   \int_0^q dq'  e^{- i E q'} \left( 
  \frac{d a}{d q'}  B - \frac{d b}{d q'} A  \right) \right]. 
\label{c20b} 
\eeq

\no 
For $\vep(x) = x$, eq.(\ref{c20b}) becomes

\beq
\psi(x)   =   \frac{1}{x^{1/2 - i E}}
\left[   C  +   \int_1^x dx'  x'^{ - i E} \left( 
  \frac{d a}{d x'}  B - \frac{d b}{d x'} A  \right) \right]. 
\label{c20c} 
\eeq

\no
The term proportional to $C$ in (\ref{c20c})
coincides with (\ref{qu9}). The integration
constants $A,B$ and $C$ are related
by eqs.(\ref{c12}) and (\ref{c13b}),

\barray 
A & = &   \frac{i E}{2} \int_0^{L_N}  dq \; 
a(q) \;  \ti(q), 
\nonumber \\
B & = &   \frac{i E}{2} \int_0^{L_N}  dq \; 
b(q) \;  \ti(q),  
\label{c21} \\
C & = & 
-  \frac{i E}{2} \int_0^{L_N} dq \; \ti(q) + 
a_1 B - b_1 A.  
\nonumber 
\earray 

\no 
Plugging (\ref{c20}) into (\ref{c21}) yields

\barray 
& (1 + R_{a,b})\,  A - R_{a,a} \,  B  - \; R_a \,  C  =   0,  
 \nonumber  &  \\
& R_{b,b} \, A + (1 - R_{b,a}) \,  B - R_b \,  C  =  0,   
\label{c22}  &  \\
& ( R_{1,b} - b_1) \, A  + ( - R_{1,a} + a_1) \, B - (R_1 + 1) \, C  =  0,&  
\nonumber 
\earray

\no
where 

\barray
R_f(E) &  = &  \frac{i E}{2} \int_0^{L_N} dq \;  f(q) \; e^{i E q},   
\label{c23} \\
 R_{f,g}(E) &  = &  \frac{i E}{2} \int_0^{L_N} dq \;  f(q) \;  e^{i E q}
  \int_0^{q} dq' \, \frac{d g}{d q'}  e^{-i E q'}.  
\nonumber 
\earray 

\no 
We need in particular

\barray
R_1(E) & = & \frac{1}{2} ( e^{i E L_N} -1 ),  
\label{c24} \\
R_{1,f}(E)  & = &   \frac{1}{2} f_1  (1 - e^{i E L_N}) - e^{i E L_N}
\tR_f(E),  \nonumber \\
R_{f,g}(E)  & = &  S_{f,g}(E)  - g_1 \; R_f(E),  \nonumber 
\earray

\no
where 

\beq
f_1 = f(x=1), \; g_1 = g(x=1), \; \tR_f(E) = R_f(-E),   
\label{c25}
\eeq

\no
and

\barray
 S_{f,g}(E) &  & = \frac{i E}{2} \int_0^{L_N} dq \;  f(q) \; g(q) 
 \label{c26} \\
& &- \frac{E^2}{2}  \int_0^{L_N} dq \;  f(q) \;  e^{i E q}
  \int_0^{q} dq' \, g(q')   e^{-i E q'}. 
\nonumber 
\earray

\no
The latter integral plays an important role in the sequel.
Eqs.(\ref{c24}) can be proved using the change in the order
of integration,

\beq 
\int_0^{L_N} d q \; \int_0^{q} d q' = 
\int_0^{L_N} d q' \; \int_{q'}^{L_N} d q.  
\label{c27}
\eeq

\no
Plugging (\ref{c24}) into (\ref{c22}) leads to

\barray 
& (1 + S_{a,b})\,  A - S_{a,a} \,  B  - \; R_a \,  
(C + b_1 A - a_1 B)   =   0, &  
 \nonumber   \\
& S_{b,b} \, A + (1 - S_{b,a}) \,  B - R_b \,  
(C + b_1 A - a_1 B)   =  0,  
\label{c28}  &  \\
&  2 e^{i E L_N} (\tR_b \, A - \tR_a B)  
+  ( e^{i E L_N} + 1)  (C + b_1 A - a_1 B)  =  0. & 
\nonumber 
\earray

\no
Observe that the term 
$C + b_1 A - a_1 B$ appears in all the equations.
This parameter determines the asymptotic
behaviour of the wave function at large values of $q$.
Indeed, the value of  $C(q)$ at $q= L_N$ (see eq.(\ref{c19})) 
is given by

\beq
C_N \equiv C(q=L_N) = C_\infty + e^{- i E L_N} ( a_N \, B - b_N \, A),  
\label{c29}
\eeq

\no
where $a_N = a(x=N), b_N = b(x=N)$,  and

\beq
C_\infty = C + A ( b_1 + 2 \tR_b) - B ( a_1 + 2 \tR_a).   
\label{c30}
\eeq

\no
In the  $\M_2$ model we shall impose

\beq
\lim_{N \rightarrow \infty} a_N = 0, \quad
\lim_{N \rightarrow \infty} b_N = 0, \quad
\label{c30b}
\eeq

\no
while for the $\M_1$ model, only the vanishing of $a_\infty$ will
be required, since $b(x) = 1$. Under these conditions, 
$C_N$ converges asymptotically to $C_\infty$, as shown
by (\ref{c29}). The latter
 parameter   characterizes
the asymptotic behaviour of the eigenfunctions,

\beq
\lim_{ q >> 1} \psi(q) \sim \frac{e^{i q E}}{\sqrt{\vep(q)}}
\;    C_\infty  + 
\dots
\label{c30c}
\eeq

\no
If $C_\infty \neq 0$, the norm of $\psi(q)$ 
behaves as $\sqrt{L_N}$ when $N \rightarrow \infty$,
and the wavefunction is delocalized, behaving
asymptotically  like the ``plane waves'' 
of the non interacting model. 
On the contrary, if  $C_\infty = 0$, the norm
of $\psi(q)$ remains finite when $N \rightarrow \infty$
for potentials decaying sufficiently fast to infinity. 
An explicit expression for the norm of these localized
states will be given at the end of section V. 
The localization
of the wave function $\psi(q)$ 
is due to an ``interference'' effect, which
can dissapear by an infinitesimal change of the potentials.
This suggest that the bound states of the model, if any, 
must be embedded in the continuum spectrum, 
unlike  ordinary QM where 
the discrete and continuum spectra usually belong
to separated regions. There are however exceptions
to this  rule, as the class of von Neumann and Wigner
oscillating potentials which have a unique bound state
with positive energy embedded in the continuum  
\cite{Galindo,neumann-wigner,simon,arai,cruz}.

Continuing with our analysis, let us combine the third
equation of (\ref{c28}) and (\ref{c30}), obtaining

\beq
C + b_1 A - a_1 B = - e^{i E L_N} \, C_\infty. 
\label{c31}
\eeq

\no
Plugging (\ref{c31}) into (\ref{c28}), yields
the following linear system of eqs.

\beq 
{\bf S} \;  {\bf w}  = 0, \quad {\bf w}^t = (A,B,C_\infty), 
\label{c32a}
\eeq

\no
where

\beq
{\bf S} =  
\left(
\begin{array}{ccc}
1 + S_{a,b} & - S_{a,a} & e^{i E L_N} R_a \\
S_{b,b} & 1 - S_{b,a} &  e^{i E L_N} R_b \\
- 2 \tR_b & 2 \tR_a &  e^{i E L_N} +1 
\end{array}
\right). 
\label{c32b}
\eeq

\no
The existence of non trivial solutions of (\ref{c32a}) requires

\beq
\det {\bf S} = 
\F(E) + \tF(E) \; e^{i E L_N} = 0,
\label{c33}
\eeq

\no
where

\beq
\F(E) = 1 + S_{a,b} - S_{b,a} + S_{a,a} \; S_{b,b} - S_{a,b} \; S_{b,a}, 
\label{c34}
\eeq

\no
and

\barray 
 \tF(E) & &  =  \F(E) 
- 2 
 \tR_a \left( R_b + R_b \; S_{a,b} - R_a \; S_{b,b} \right)
\nonumber \\
& &
 + 2  \tR_b \left(  
 R_a - R_a \; S_{b,a} + R_b \; S_{a,a} \right).  
\label{c35} 
\earray

\no
To simplify $\tF(E)$, we use the equation

\beq
S_{f,g}(E) + S_{g,f}(-E) = - 2 R_f(E) \; R_g(-E),  
\label{c37}
\eeq

\no
which implies

\beq
 \tF(E) =  \F(-E). 
\label{c36}
\eeq

\no
The final form of the eigenenergies equation is

\beq
\F(E) + \F(-E) \; e^{i E L_N} = 0.
\label{c38}
\eeq

\no
Before we analyze in detail the possible solutions 
of (\ref{c38}) we shall make some comments.

\begin{itemize}

\item In the absence of interactions, i.e. $a(x) = b(x) = 0$, 
one gets $\F(E) = 1 ,\; \forall E$,  and then eq.(\ref{c38}), 
for the case $\vep(x) = x$, 
reproduces eq.(\ref{qu20}).

\item If $E$ is a solution of (\ref{c38}), so is $-E$
for generic potentials $a$ and $b$, including 
complex functions. This is  a consequence
of the antisymmetry of $H$.

\item $\F(E)$ is invariant under the  
$SL(2,\R)$ transformation (\ref{c3}). In fact, the
terms $S_{a,b} - S_{b,a}$ and  $S_{a,a} \; S_{b,b} - S_{a,b} \; S_{b,a}$ 
are invariant separately.

\item We shall assume that the integrals defining $R_f(E)$ and $S_{f,g}(E)$, 
converge for all values of $E$. At $E=0$ this implies

\beq
R_f(0) =  S_{f,g}(0) = 0, \; \;  f, g = a, b  \Longrightarrow \F(0) = 1,  
\label{c40}
\eeq

\no
Hence $E=0$ is not a solution of  eq.(\ref{c38})
and therefore the Hamiltonian $H^{-1}$ is non singular
as  assumed at the beginning of this section.

\end{itemize}

Returning to the solution of eq.(\ref{c32a}), let us define 
the 3-component vectors,

\barray 
{\bf v}_1 & = & (1 + S_{a,b},\; - S_{a,a}, \;   e^{i E L_N} \, R_a), 
\nonumber  \\
{\bf v}_2  & = &  (S_{b,b},\;  1 - S_{b,a}, \;  e^{i E L_N}\, R_b),
\label{c41}
 \\
{\bf v}_3 & =  & (- 2 \tR_b ,\;  2 \tR_a, \;  e^{i E L_N} +1), 
\nonumber  
\earray 

\no
which are the rows of the matrix ${\bf S}$, and  are
linearly dependent by (\ref{c33}). These vectors
span a plane which can be characterized by its normal ${\bf n}$.
The vector ${\bf w}$ that  that solves (\ref{c32a}),
must be proportional to ${\bf n}$.   
If  ${\bf v}_1$ and ${\bf v}_2$ are non collinear, we can choose

\beq
{\bf w} = {\bf v}_1 \times {\bf v}_2, \qquad
{\bf v}_1 \nparallel {\bf v}_2,  
\label{c42}
\eeq

\no
where

\barray 
&  {\bf v}_1 \times {\bf v}_2 =  &  \label{c43} \\
 & ( - e^{i E L_N}  (  R_b S_{a,a}+   R_a ( 1 - S_{b,a})), & 
\nonumber  \\
&  e^{i E L_N} ( R_a S_{b,b} - R_b (1 + S_{a,b})), \, \F). & 
\nonumber 
\earray

\no
We get in particular

\beq
C_\infty = \F(E).  
\label{c44}
\eeq

\no
Hence the delocalized eigenstates, i.e. $C_\infty \neq 0$, 
satisfy that $\F(E) \neq  0$, 
while the localized ones,  i.e. $C_\infty = 0$, 
correspond to  $\F(E)=0$. 

Otherwise, if ${\bf v}_1$ and  ${\bf v}_2$ 
are collinear, the vector ${\bf w}$ can  be choosen as

\beq
{\bf w} = {\bf v}_1 \times {\bf v}_3, \qquad
{\rm if} \;\; 
{\bf v}_1 \parallel {\bf v}_2, \; {\bf v}_1 \nparallel {\bf v}_3. 
\label{c45}
\eeq

\no
This case is exceptional since
it requires the vanishing of the 
functions giving ${\bf v}_1 \times {\bf v}_2$.
Another unlikely  possibility
is that the three vectors ${\bf v}_i$ are
collinear, in which case the vector ${\bf w}$ 
must belong to the plane orthogonal to  ${\bf v}_{1,2,3}$.

\begin{figure}[t!]
\begin{center}
\includegraphics[height= 1.7 cm,angle= 0]{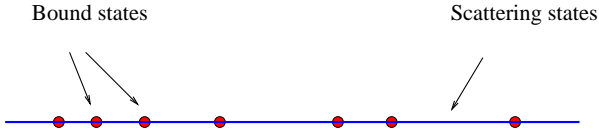}
\end{center}
\caption{
Pictorial representation of the spectrum of the model.
The bound states are the points where $\F(E) = 0$, which
are embedded in a continuum of scattering states. 
}
\label{bound-states}
\end{figure}

In summary, the most common situation we shall
encounter is described by eqs.(\ref{c42}),
so that the localized/delocalized nature of 
the eigenfunctions is fully characterized by the 
vanishing/non vanishing of $\F(E)$. 
The generic structure of the spectrum is depicted 
in figure \ref{bound-states} and table 3.

In so far we have not used the reality of the potentials
$a$ and $b$, which guarantees the hermiticity of the
Hamiltonian (\ref{c1}). If they are real functions, then $\F(E)$ is 
a complex hermitean function, i.e.

\beq
a(x), b(x) \in \R \Longrightarrow  \F^*(E) = \F(-E^*). 
\label{c46}
\eeq

\no
This equation follows from the identity

\beq
S_{f,g}^*(E) = S_{f^*, g^*}(-E^*).  
\label{c47}
\eeq

\no
If $E$ is real, eq.(\ref{c46}) implies that $\F^*(E) = \F(-E)$  
and hence the eigenvalue equation (\ref{c38}), for
delocalized states, can be written as

\beq
e^{i E L_N} = - \frac{\F(E)}{\F(E)^*},\qquad {\rm for} \; \F(E) \neq 0
\label{c48}
\eeq

\no
The RHS of this equation describes the scattering phase
shift produced by the interaction. For  localized
states, equation (\ref{c48}) becomes singular, 
but eq.(\ref{c38}) is automatically satisfied
since $\F(E) = \F(-E)= 0$.
All these results means that  $\F(E)$ 
plays, in our model, 
the role of a Jost function which determines
completely the scattering phases and bound states. 
However there are some important differences concerning their
analytical properties that we shall discuss below.

\begin{center}
\begin{tabular}{|c|c|c|c|}
\hline
Eigenstate & $C_\infty$  &  $\F(E)$ &  Eigencondition \\
\hline 
Delocalized & $\neq 0$ & $\neq 0$ & $e^{i E L_N} = -
\frac{\F(E)}{\F(E)^*} $ \\
Localized & $= 0$ & $ = 0$ & $\F(E) = 0$ \\
\hline
\end{tabular}

\vspace{0.5 cm}

Table 3.- Classification of eigenstates of the $\M_2$ model.
For the $\M_1$ model,  $\F(E)$ is replaced by $\F_1(E)$. 
\end{center}

\subsection{Schr\"odinger equation for the $\M_1$ model}

The diagonalization of the Hamiltonian $H_1$ (\ref{c2})
proceeds along the same steps as for $H_2$ with the suitable 
changes. The main difference is that the phase factor
$e^{i E L_N}$ arises in several expressions which 
one needs to extract out and factorize conveniently. 
In particular, from (\ref{c26}) one finds 

\barray
& S_{a,1}   =   R_a , \;   S_{1,a}  =   - e^{i E L_N} \; \tR_a,
\;  S_{1,1} = \frac{1}{2} (e^{i E L_N} -1),   & 
\nonumber  
\earray

\no
and then

\beq
 \F  =  1 + R_a - \frac{1}{2} S_{a,a} + e^{i E L_N} \left(
\tR_a +   \frac{1}{2} S_{a,a} + R_a \; \tR_a \right),
\label{c50}
\eeq

\no
so that eq.(\ref{c38}) becomes

\beq
 \F_1(E) + e^{i E L_N} \,  \tF_1(E) = 0,   
\label{c51}
\eeq

\no
where

\beq
\F_1(E) = 1 + 2 R_a - S_{a,a}, \; 
\tF_1(E) = 1 + 2 \tR_a + S_{a,a} + 2 R_a \; \tR_a.  
\label{c52}
\eeq

\no 
Using (\ref{c37}) for $f=g=a$, one finds the  
 analogue of (\ref{c36}),

\beq
 \tF_1(E) =  \F_1(-E), 
\label{c53}
\eeq

\no
so that

\beq
 \F_1(E) + e^{i E L_N} \,  \F_1(-E) = 0.   
\label{c54}
\eeq

\no
For the $\M_1$ model the constant $C_\infty$ 
is no longer related to the asymptotic value 
of $C_N$ (recall eqs.(\ref{c29}) and (\ref{c30})). 
Instead, it is replaced by the parameter $C_{1,\infty}$
which satisfies

\barray 
C_N & = & C_{1, \infty} + e^{- i E L_N} \, a_N B, 
\label{c55} \\
C_{1,\infty} & = & C_\infty - e^{- i E L_N } A, 
\\
C_{1, \infty} & = & - e^{i E L_N} ( C + 2 A - a_1 B),  
 \nonumber
\earray

\no
where we have used eq.(\ref{c31}).  The linear system
(\ref{c28}) for the constants $A,B, C_{1,\infty}$
turns into

\beq 
{\bf S}_1 \;  {\bf w}_1  = 0,  \quad {\bf w}_1^t = (A,B,C_{1,\infty}), 
\label{c56}
\eeq

\no
where

\beq
{\bf S}_1 =  
\left(
\begin{array}{ccc}
1 + 2 R_a & - S_{a,a} & e^{i E L_N} R_a \\
-1 & 1  & - e^{i E L_N}  \\
2 & 2 \tR_a &  e^{i E L_N} +1 
\end{array}
\right), 
\label{c57}
\eeq

\no
and whose determinant reproduces eq.(\ref{c54}),

\beq
\det {\bf S}_1 = 
\F_1(E) + \tF_1(E) \; e^{i E L_N}.  
\label{c58}
\eeq

\no
Repeating the analysis made for the $\M_2$ model
we obtain the analogue of eq.(\ref{c44}),

\beq
C_{1,\infty} = \F_1(E),  
\label{c59}
\eeq

\no
so that  $\F_1(E)$
controls the delocalized/localized character of the eigenfunctions.
The unique exceptional case appears when $\F_1(E) = 0$, $R_a(E) = -1$ and
$e^{ i E L_N} = 1$ where, besides a localized solution with
$(A,B,C_{1,\infty}) = (1,1,0)$, there is the delocalized one
$(A,B,C_{1,\infty}) = (1,-1,-2)$. In summary, the generic
eigenstates of the $\M_1$ model 
are described by table 3 with $C_\infty$ and
$\F(E)$ being replaced by  $C_{1,\infty}$ and $\F_1(E)$.
Most of the comments  concerning eq.(\ref{c38})
also apply to (\ref{c53}). In  addition we add:

\begin{itemize}

\item The function $\F_1$, unlike $\F$, 
is not manifestly invariant under the transformation 
(\ref{c5}), under which

\beq
\F_1 \rightarrow \F_1 + \alpha \left[
e^{i E L_n} (\tR_a +1) - R_a -1 \right] - \alpha^2
(e^{i E L_n} -1). 
\label{c60}
\eeq

\no
However, eq.(\ref{c54}) is invariant under this change,
as can be easily proved. In the large $N$ limit
we shall impose that $a_N \rightarrow 0$, hence
the symmetry (\ref{c5}) will be fixed.

\item  Eq.(\ref{c37}) implies that

\beq
\F_1(E) + \F_1(-E) = 2 ( 1+ R_a(E)) (1 + R_a(-E)). 
\label{c61}
\eeq

\no
If $a(x)$ is real, the LHS of (\ref{c61}) gives the real
part of $\F_1(E)$, while the RHS is the norm squared
of the function $f_1(E)$ (up to some constants)

\beq
\Re \;  \F_1(E) = | f_1(E)|^2 \geq 0, \qquad 
f_1(E) =  1+ R_a(E).   
\label{c63}
\eeq

\no 
Hence for the $\M_1$ model, the real part
of $\F_1(E)$ is always positive. 
This property is analogue to the positivity 
of the imaginary part of the scattering amplitudes
in some many body and QM models. On the contrary
the real part of $\F$ can be positive or 
negative as we shall show in an example below.

\end{itemize}

\subsection{The zeros of the Jost functions}

The Jost functions  $\F(E)$ and $\F_1(E)$
depend in general on the system size $N$. To make this dependence
explicit, we shall denote them by $\F_N(E)$ and $\F_{1,N}(E)$. 
We shall assume that these functions are well
defined in the limit $N \rightarrow \infty$, and call 
them   $\F_\infty(E)$ and $\F_{1,\infty}(E)$. 

The reality of 
 $a(x)$ and $b(x)$ guarantees the hermiticity of the Hamiltonians 
$H_{1,2}$, i.e. 

\beq
H_{1,2}^\dagger = H_{1,2} \Longleftrightarrow
a(x)^* = a(x), \quad b(x)^* = b(x), \;\; \forall x,
\label{c64}
\eeq

\no  
and in turn that of the spectrum. 
Under these conditions, we shall prove the following 
important result:

\beq
{\rm if} \; \F_\infty(E) = 0  
 \Longrightarrow \Im \, E \leq 0.  
\label{c65}
\eeq

\no 
A similar statement holds for  $\F_{1,\infty}(E)$. 
In other words,  the zeros of the Jost functions,
at $N = \infty$, always lie either 
 on the real axis, corresponding to localized states, or
below it, corresponding to resonances.  
The proof of (\ref{c65}) is straightforward. 
Suppose that $E$
satisfies eq. (\ref{c38}), which implies that $E$ is an eigenvalue
of the Hamiltonian $H_2$. Since the latter is hermitean, 
$E$ must be a real number, i.e.

\beq
{\rm if} \;  \F_N(E) + \F_N(-E) \; e^{i E L_N} = 0
\Longrightarrow  \Im \, E = 0.
\label{c66}
\eeq

\no
Negating this implication yields,

\beq
\forall E, L_N  \;\; {\rm if} \;   \Im E \neq 0 
\Longrightarrow   \F_N(E) + \F_N(-E) \; e^{i E L_N} \neq 0.
\label{c67}
\eeq

\no
Restricting to the case where $\Im \, E  > 0$ and taking the limit
$N \rightarrow \infty$ (i.e. $L_N \rightarrow \infty)$ in (\ref{c67})
one gets

\beq
\forall E, \; {\rm if}  \; \; \Im E > 0 
\Longrightarrow   \F_\infty(E) \neq 0, 
\label{c68}
\eeq

\no
where the factor $e^{i E L_N}$ converges toward zero 
and thus cancells out the second term in (\ref{c67}). 
Negating eq.(\ref{c68}) yields the desired statement
(\ref{c65}). The case of 
 $\F_{1,\infty}(E)$ is similar. Repeating 
this argument in the case where 
$\Im \, E < 0$ does not give further information
on the zeros of $\F(E)$. 

This property of the zeros of the Jost functions $\F(E)$ and
$\F_1(E)$ is  quite remarkable. In standard QM 
the zeros of the Jost function appear in the 
imaginary axis of the complex momentum plane (corresponding to
bound states), or below the  
real axis (corresponding to resonances).  
Eq.(\ref{c65})  suggests a way
to prove the Riemann hypothesis. Suppose, for a while, that
$\F(E)$ were proportional  to $\zeta(1/2 - i E)$. Hence, 
since the $\F(E)$ cannot have zeros with $\Im E > 0$,
the same property holds for   $\zeta(1/2 - i E)$.
The latter statement implies  
the RH. In section V, we shall see that   $\zeta(1/2 - i E)$
does not have the correct analyticity 
properties to become a Jost function
of the model, but some modification of it may
in principle. 
This issue will be considered in section VI.

\subsection{Examples of Jost functions}

To illustrate the general 
solution of the Hamiltonians $H_{1,2}$ we shall
consider some simple models.

\subsubsection*{Example 1:  step potential in the  $\M_1$ model}

Let us take 

\beq
a(x) = a_1 \; \theta(x_1 - x), \qquad 1 \leq x \leq N,    
\label{ex1}
\eeq

\no
where  $\theta(x)$ is the Heaviside step function
and $1 < x_1 < N$. $R_a$ and $S_{a,a}$
are readily computed from eqs.(\ref{c23})
and (\ref{c26}),

\beq
R_a =   \frac{a_1}{2} \;( e^{i q_1 E } -1), \qquad  
S_{a,a}  =    \frac{a_1^2}{2} \;( e^{i q_1 E } -1), 
\label{ex2}
\eeq

\no
where $q_1 = \log x_1$. The associated Jost function
$\F_1(E)$ follows from (\ref{c52}),

\beq
\F_1(E) = 1 + \frac{a_1 \;( 2 - a_1)}{2}  ( e^{i q_1 E} -1). 
\label{ex3}
\eeq

\begin{figure}[t!]
\begin{center}
\includegraphics[height= 4 cm,angle= 0]{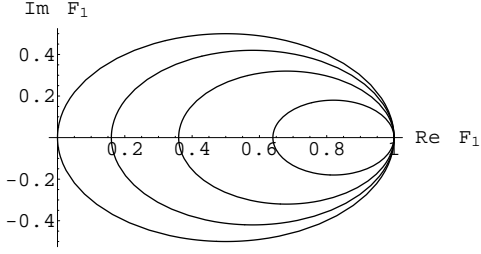}
\end{center}
\caption{
Plot of the real and imaginary parts of $\F_1(E)$,
as given by eq.(\ref{ex3}), for the choices
$a_1 = 0.2, 0.4, 0.6, 1$. At $a_1= 1$ the circle
passes through the origin. 
}
\label{step1}
\end{figure}

\no 
For each value of $a_1 \neq 0, 2$, the real and imaginary parts 
of $ \F_1(E)$ describe a 
circle  (see fig. \ref{step1}).
At $a_1 = 1$ the circle touches the origin, i.e. $\F_1 =0$,
at the energies

\beq
e^{i q_1 E} = -1 \Longrightarrow 
E^{(I)}_{n_1} = \frac{(2 n_1 + 1) \pi}{q_1},\; \; n_1 = 0, \pm 1, \dots 
\label{ex4}
\eeq

\no
describing an infinite number of bound states. 
The eigenstates corresponding to
$\F_1 \neq 0$ satisfy

\beq
  e^{i E  L_N }   = - 
 \frac{\F_1}{\F_1^*} = - \;  e^{i q_1 E}, 
\label{ex5}
\eeq

\no
and their energy is

\beq
E^{(II)}_{n_2} = \frac{(2 n_{2} + 1) \pi}{L_N - q_1}, \; \; 
n_{2} = 0, 
\pm 1, \dots
\label{ex6}
\eeq

\no
In the thermodynamic limit, where  $L_N \rightarrow \infty$ 
and $q_1$ is kept fixed, the spectrum of the Hamiltonian
at $a_1=1$ consists of a continuum formed by the 
eigenenergies (\ref{ex6}), and a discrete part formed by 
(\ref{ex4}) (see fig. \ref{bound-states}). The explanation of these results
is straightforward. The Hamiltonian $H_1^{-1}$, for the potential
(\ref{ex1}), has the block diagonal form

\beq
H_1^{-1}(x, x')  =  \frac{i}{ 2  \sqrt{\vep(x)\vep(x')}}   
\left( 
\begin{array}{cc|cc}
0 & - {\bf 1} & a_1-1 & a_1 -1  \\
{\bf 1} & 0 &   a_1-1 & a_1 -1 \\
\hline
1- a_1 & 1 - a_1 & 0 & - {\bf 1} \\
1- a_1 & 1 - a_1 & {\bf 1} & 0 \\
\end{array}
\right), 
\label{ex6b}
\eeq

\no
where the vertical and horizontal lines separate the regions 
$1 < x < x_1$ and $x_1 < x < N$, 
and ${\bf 1}$ denotes a matrix with all entries
equal to 1. At $a_1 = 1$, the matrix (\ref{ex6b})
splits into two commuting blocks whose structure
is identical  to that of $H_0^{-1}$ in the corresponding 
intervals. 
Eqs.(\ref{ex4}) and (\ref{ex6}) simply correspond
to  the non interacting eigenenergies in those  regions. 
For $a_1 \neq 1$,  the Hamiltonian
$H_{1}^{-1}$ is non diagonal and the eigenstates are
plane waves in all the regions, 
with a phase factor $e^{i q E}$ and 
a discontinuity in the amplitude at $x = x_1$. 

The function (\ref{ex3}) is invariant under the transformation

\beq
a_1 \leftrightarrow 2 - a_1, 
\label{ex7}
\eeq

\no
which maps $a_1 = 0$ into $a_1 = 2$, and has $a_1=1$ as a fixed
point. At $a_1=2$ one has $\F = 1, \; \forall E$, so that  
all the energy levels satisfy $e^{i  E L_N} = -1$. 
One can check that the  function $\ti$ is a plane
wave $e^{i q E}$, which changes its sign after crossing
$q=q_1$. More generally, one can define a unitary transformation
which changes the sign of $\psi$ for $q > q_1$. Under
this transformation the Hamiltonian for $a_1$ is mapped
into that of $2-a_1$, which explains eq.(\ref{ex7}).

Finally, let us look for the zeros of $\F_1(E)$ for generic
values of $a_1$,

\beq
\F_1(E) = 0 \Longrightarrow e^{i E q_1} 
= \frac{ (a_1 -1)^2 + 1}{(a_1 - 1)^2 -1}.
\label{ex8}
\eeq

\no
For $a_1$ real, the RHS of (\ref{ex8}) has an  absolute value
greater than one, and therefore the imaginary part of $E$ is a
non positive number,

\beq
\F_1(E) = 0 \; {\rm and} \; 
a_1 \in \Rmath  \Longrightarrow
\Im \, E \leq 0,  
\label{ex9}
\eeq

\no
in agreement with  eq.(\ref{c65}).

\subsubsection*{Example 2:  step potentials in the  $\M_2$ model}

Let us take

\beq
a(x)  =   a_1 \; \theta(x_1 - x), \quad 
b(x)  =   b_1 \; \theta(x_2 - x), \qquad 1 \leq x \leq N,    
\label{ex10} 
\eeq

\no
where  $1 < x_1 < x_2 < N$. The $S$-functions are given by

\barray 
S_{a,a}  =   \frac{a_1^2}{2} \;( e^{i q_1 E } -1), & &  
S_{b,b}  =   \frac{b_1^2}{2} \;( e^{i q_2 E } -1),  
\label{ex11}
 \\
S_{a,b}  =  \frac{a_1 b_1}{2} \;( e^{i q_1 E } -1), & &  
S_{b,a}  =   \frac{a_1 b_1}{2} \; e^{i (q_2 - q_1) E} ( e^{i q_1 E } -1) 
\nonumber 
\earray

\no
where $q_i = \log x_i \; (i=1,2)$. The Jost function
is given by

\beq
\F(E) = 1 + c ( c-1)   ( e^{i q_1 E} -1)
( e^{i (q_2 - q_1) E} -1), \; \; \; \;  c = \frac{a_1 b_1}{2}.  
\label{ex12}
\eeq

\no
The conditions for this function to vanish, for real values
of $E$, are

\beq
c = \frac{1}{2}, \quad 
\frac{q_1}{q_2} = \frac{4 n_1 \pm 1 }{4 n_2 + 1 \pm 1 }, \;\;
\; n_1, n_2 \in \Nmath, 
\label{ex13}
\eeq

\no
which gives the eigenenergies

\beq
E_{n_1,n_2} = \frac{\pi}{2 q_1} (4 n_1 \pm  1)=
   \frac{\pi}{2 q_2} (4 n_2 + 1 \pm  1). 
\label{ex14}
\eeq

\begin{figure}[t!]
\begin{center}
\includegraphics[height= 4 cm,angle= 0]{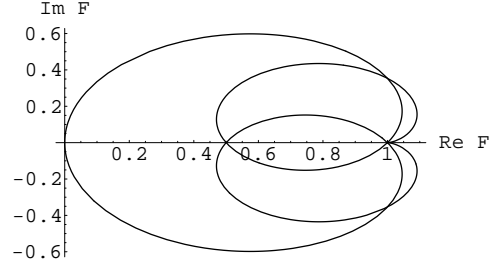}
\end{center}
\caption{
Plot of the real and imaginary parts of $\F(E)$
as given by eq.(\ref{ex12}) for the choice
$c= 1/2$ and $q_1/q_2 = 3/4$ in eq.(\ref{ex13}). 
}
\label{step2}
\end{figure}

\no 
Fig.\ref{step2} shows a particular example. 
To  check  that the zeros of $\F(E)$
lie below the real axis 
for generic real values of $q_1, q_2$ and $c$ one 
writes the equation $\F(E) = 0$ as

\beq 
e^{ i (q_2 - q_1) E} = 1 - \frac{1}{c ( c-1) ( e^{i q_1 E} -1)}.
\label{ex15}
\eeq

\no
If $\Im \, E  > 0$, the LHS of this equation is a complex
number with modulus less than one, in contradiction with the
fact that the RHS has  modulus
greater than one. Hence one must have  $\Im \, E  \leq 0$.

\subsubsection*{Example 3:  algebraic potential in the  $\M_1$ model}

Let us choose

\beq
a(x) = \frac{a_1}{x^\mu}, \quad \mu  > 0,   \qquad 1 \leq x \leq \infty. 
\label{ex16}
\eeq

\no 
The $R$ and $S$ functions are given by

\beq
R_a = - \frac{a_1}{2} \frac{E}{E + i \mu}, \quad
S_{a,a} =  - \frac{a_1^2}{4} \frac{E}{E + i \mu},
\label{ex17}
\eeq

\no
so that

\beq
\F_1(E) = 1 + \left( \frac{a_1^2}{4}  - a_1 \right) \frac{E}{E + i \mu}.
\label{ex18}
\eeq

\no
This  function vanishes at

\beq
\F_1(E) = 0 \Longrightarrow E = - i \frac{ \mu}{ (1 - \frac{a_1}{2})^2}, 
\label{ex19}
\eeq

\no
in agreement with (\ref{c65}), and it also has a pole at

\beq
\F_1(E) = \infty \Longrightarrow E = - i  \mu,
\label{ex19-b}
\eeq

\no
which belongs to the lower half plane. The latter property
is a general feature of the Jost functions, which
consist in that their singularities always lie below
the real axis. This result will be proved in the next section.

\subsubsection*{Example 4:  algebraic potentials in the  $\M_2$ model}

Let us choose

\beq
a(x) =  \frac{a_1}{x^{\mu_1}}, \quad 
b(x) =  \frac{b_1}{x^{\mu_2}},  \quad 
\mu_{1,2}  > 0,  \;\;  1 \leq x \leq \infty,  
\label{ex20} 
\eeq

\no
with $\mu_1 \neq \mu_2$. The $S$-functions are given by

\barray 
S_{a,a} =  - \frac{a_1^2}{4} \frac{E}{E + i \mu_1}, & & 
S_{b,b} =  - \frac{b_1^2}{4} \frac{E}{E + i \mu_2},
\label{ex21}
 \\
S_{a,b} =  - \frac{a_1 b_1 \mu_1}{2 ( \mu_1 + \mu_2)} 
\frac{E}{E + i \mu_1}, & & 
S_{b,a} =  - \frac{a_1 b_1 \mu_2}{2 ( \mu_1 + \mu_2)} 
\frac{E}{E + i \mu_2}, 
\nonumber
\earray

\no
which yields

\beq
\F(E) = 1 + \left(\frac{\rho^2}{4} - \rho \right) 
\frac{E^2}{(E + i \mu_1)(E + i \mu_2)},
\label{ex22}
\eeq

\beq
\rho = \frac{ a_1 b_1 ( \mu_1 - \mu_2)}{2 ( \mu_1 + \mu_2)}. 
\label{ex22-b}
\eeq

\no
One can easily check that the zeros and the poles
of this function always lie below the real axis. 

We have investigated other potentials 
to check explicitely the property (\ref{c65}). 
For  steps potentials, where the $q$ intervals
are related by rational fractions, 
eq.(\ref{c65})  follows 
from the Routh-Hurwitz theorem for the 
localization of the zeros of polynomials with real coefficients
\cite{GR}. For more general potentials we have been able to
check (\ref{c65})  numerically but not analytically. 
These results suggest the $\M_{1,2}$ models
may provide a huge class of complex functions 
with that interesting property.

\section{Analiticity properties of the Jost functions}

In Quantum Mechanics the Jost function
display analiticity
properties which are a consequence of causality. 
The close link between causality and analiticity 
is illustrated by the following theorem due
to Titchmarsh \cite{Titchmarsh,AW}. 

Let $f(q)$ be a generic complex function and  $\hat{f}(E)$
 its Fourier transform,

\beq
\hat{f}(E) = \int_{- \infty}^\infty dq \, f(q) \, e^{i E q}, \quad
 f(q) = \int_{- \infty}^\infty \frac{dE}{2 \pi} \, \hat{f}(E) \, e^{-i E q}.
\label{j1}
\eeq

\no
Assuming that  $\hat{f}(E)$ is  square integrable
over the reals axis,

\beq
 \int_{- \infty}^\infty dE \,| \hat{f}(E)|^2 \; < \infty,
\label{j2}
\eeq

\no
then, any of the following three statements implies
the other two:

\begin{itemize}

\item  1) $f(q) = 0$ for $q < 0$.

\item 2) $\hat{f}(z)$ is analytic in the upper half plane, 
$\Im \, z > 0$, and approaches $\hat{f}(x)$  almost everywhere 
as $y \rightarrow 0$. Further

\beq
 \int_{- \infty}^\infty dx \,| \hat{f}(x + i y)|^2 \; < K, \;\; y > 0.  
\label{j3}
\eeq

\item 3) The real and imaginary parts of  $\hat{f}$, 
on the real axis, are the Hilbert transforms of each other,

\barray 
& u = \H[v], \; 
v = - \H[u], \; 
\hat{f}(x) =  u(x) + i v(x),  
& 
\label{j4} 
\earray

\beq
 \H[g](x) = P \int_{- \infty}^\infty \frac{dy}{\pi} 
\frac{g(y)}{y -x},  
\label{j4b}
\eeq

\no 
where $P$ denotes the Cauchy principal value of the integral.

\end{itemize}

Statement 1) is called causality, which in the present
context means that the functions
$a(q)$ and $b(q)$ are zero for negative $q$-times, i.e. 

\beq
a(q) = b(q) = 0 \quad {\rm for} \; q < 0.  
\label{j5}
\eeq

\no
In fact, the variable $q$ is always non negative
by eq.(\ref{c15}), so that (\ref{j5}) 
must be understood as an extension of the
definition of $a(q)$ and $b(q)$ for negative 
values of $q$.
The main consequence of causality
are the dispersion relations (\ref{j4}),
which play a central role in 
the scattering theory in Quantum Mechanics, 
and other fields of Physics. 

For the $\M_2$ model to be well
defined in the limit  $N \rightarrow \infty$,
we shall impose that $f = a,b$ are 
 square integrable functions, i.e.

\beq
 \int_{-\infty}^\infty dq \,|f(q)|^2 \;  = 
 \int_{-\infty}^\infty  \frac{dE}{2 \pi}  \,|\hat{f}(E)|^2 \;  < \infty.  
\label{j6}
\eeq

\no
Hence  by the Titchmarsh theorem,  
$\hat{a}(E)$ and $\hat{b}(E)$ 
are analytic  functions in the upper half plane
and satisfy eq.(\ref{j4}), which can be combined into

\beq
\hat{f}(E) =  
\int_{- \infty}^\infty \frac{dt}{i \pi} \;\;  \frac{\hat{f}(t)}{t - E},
\quad f = a,b.  
\label{j7}
\eeq

\no
These properties, in turn, imply that  $S_{f,g}$ and $\F$
are analytic functions.

\subsection{Analyticity of $S_{f,g}$}

Consider the $S$-function defined in (\ref{c26}) in the limit
$N \rightarrow \infty$,

\barray 
S_{f,g}(E)&  = &   \frac{i E}{2} \int_0^{\infty} dq \;  f(q) \; g(q)
\label{j8} \\
& &  - 
 \frac{E^2}{2} \int_0^{\infty} dq \;  f(q) \;  e^{i E q}
  \int_0^{q} dq' \, g(q')   e^{-i E q'},
\nonumber 
\earray

\no
where $f$ and $g$ are  causal functions, in the sense
of (\ref{j5}), and square normalizable. 
Replacing $f$ and $g$ by their
Fourier transforms one arrives at

\barray 
S_{f,g}(E) &  =  &  \frac{i E}{2} \int_{- \infty}^{\infty} \frac{dt}{2 \pi}
 \;  \hat{f}(t) \; \hat{g}(-t) 
\label{j9} \\
& & - 
 \frac{i E^2}{2} \int_{- \infty}^{\infty}  \frac{dt}{2 \pi} \; 
\frac{ \hat{g}(t)}{ t + E} \left( \hat{f}(-t) - \hat{f}(E) \right).  
\nonumber
\earray

\no
Using (\ref{j7}) for $\hat{g}$ yields,

\beq
S_{f,g}(E) = - \frac{E^2}{4}  \hat{f}(E) \; \hat{g}(-E) + 
 \frac{i E}{4} \int_{- \infty}^{\infty}  \frac{dt}{ \pi} \; 
\frac{ t \;  \hat{f}(t)  \; \hat{g}(-t) }{ t - E},   
\label{j10}
\eeq

\no
or equivalently

\barray 
s_{f,g}(E)&  \equiv &  - \frac{4 \; S_{f,g}(E)}{E} 
\label{j11} \\
& & 
 = E \hat{f}(E) 
\; \hat{g}(-E) - i 
 \int_{- \infty}^{\infty}  \frac{dt}{ \pi} \; 
\frac{ t \;  \hat{f}(t)  \; \hat{g}(-t) }{ t - E}.   
\nonumber 
\earray 

\no 
which shows that  $s_{f,g}(E)$
 is the sum of the function  $k(E) = t f(E) \, g(-E)$ 
and its Hilbert transform, i.e.

\beq
s_{f,g} = k - i \H[k]. 
\label{j12}
\eeq

\no
If $k(t)$ belongs to the space $ L_p \equiv L_p(-\infty, \infty)$, then 
$\H[k]$ is defined and belongs to  $L_p$
 for $p > 1$ \cite{enciclopedia}. 
In this case the Hilbert transform
of $s_{f,g}$ satisfies eq.(\ref{j7}). Moreover, if
$s_{f,g}(E)$ is square normalizable then,
by the Titchmarsh theorem, 
it will be an analytic 
function in the upper half plane, i.e.

\beq
{\rm If} \; \;   t  f(t)  g(-t) \in 
L_{2} \Longrightarrow
s_{f,g} : {\rm analytic} \; {\rm in} \;  \Cmath_+
\label{j13}
\eeq

\no
Apparently the normalizability of $f$ and $g$ does
not guarantee that of $s_{f,g}$, but it 
all the examples we have analyzed that is the case.

\subsection{Analyticity of $\F_1(E)$}

A consequence of eq.(\ref{j13})  is

\beq
{\rm if} \; \;  t \, |\hat{a}(t)|^2 \in 
L_{2}  \Longrightarrow
\frac{\F_1(E) - 1}{E} : {\rm analytic} \; {\rm in} \;  \Cmath_+. 
\label{j14}
\eeq

\no
To prove (\ref{j14}), write  $\F_1(E)$ as (recall eqs. (\ref{c52})
and (\ref{c23}))

\beq
\F_1(E) = 1 + i E \;  \hat{a}(E) - S_{a,a}(E).   
\label{j15}
\eeq

\no
Recall that  $\hat{a}$ is analytic in $\Cmath_+$ as
well as $S_{a,a}(t)/t$, provided that $t \;  |\hat{a}(t)|^2 \in L_{p > 1}$. 
Hence (\ref{j14}) follows. 
We also expect that under appropiate conditions
on the potentials, 
the combination $(\F(E) -1)/E$ will be analytic
in the upper half plane.

We proved in section IV that the real part of 
$\F_1(E)$ is always positive and equal to the square
of the function $f_1(E)$ (see (\ref{c63})),

\beq
f_1(E) = 1 + \frac{i E}{2}  \;  \hat{a}(E).  
\label{j16}
\eeq

\no
Using eq.(\ref{j15}) and the analyticity of $\hat{a}$ one
can prove that the imaginary part of $\F_1(E)$ is given
by the Hilbert transform of  $|f_1(E)|^2$, i.e.

\beq
\F_1(E) = |f_1(E)|^2 - i P  \int_{- \infty}^\infty 
\frac{dt}{ \pi} \; \frac{  |f_1(t)|^2}{t - E}. 
\label{j17}
\eeq

\no 
However, the converse is not true. The reason is that
$\F_1(E)$ does not in general converge toward zero as
${\rm Im} \;  E$ goes to infinity.  
This can be simply illustrated by the non-interacting case
where  $\F_1(E) = 1$. 

Finally, we shall give the expression of 
the localized eigenstates of the $\M_2$ model
in the limit $L_N \rightarrow \infty$, i.e.

\barray 
\langle \psi_E | \psi_E \rangle &  = &  \int_0^\infty
dq \; | \ti(q) |^2 \label{j18} \\
& & 
=
(A^*, B^*) 
\left( 
\begin{array}{cc}
\Omega_{b,b} & - \Omega_{a,b} \\
- \Omega_{b,a} & \Omega_{a,a} \\
\end{array}
\right) 
\left( 
\begin{array}{c}
A \\
B \\
\end{array}
\right), 
\nonumber 
\earray

\no
where

\beq
\Omega_{f,g} =  - 2 \left(
S_{g, q f} + \tilde{S}_{f, q g} 
+ \frac{i}{E} ( 
S_{g,f} - \tilde{S}_{f,g} ) \right) - \int_0^\infty
dq \;  f g. 
\label{j19}
\eeq

\no
for $f,g = a,b$. Using the Fourier transforms of these
functions one can write (\ref{j19}) as

\barray 
& & \Omega_{f,g}(E) =  \label{j20} \\
& & 
 P \; \int_{- \infty}^\infty
\frac{dt}{2 \pi} f(t)^*  \left[
\frac{t+E}{t - E}   + 
\frac{t \, E}{t - E}   
\left( \frac{\overrightarrow{d} }{dt} + 
 \frac{\overleftarrow{d} }{dt} \right) 
 \right] g(t)  \nonumber \\
& & 
- 
\frac{i E^2}{2} 
  f(E)^\*  
\left( \frac{\overrightarrow{d} }{dE} - 
 \frac{\overleftarrow{d} }{dE} \right) g(E)
\nonumber \\
\earray

\no
where $\overleftarrow{d}/dt$ only acts on the 
function  $f(t)^*$. 
The results obtained on this section can be
given a more formal treatment using the 
 Theory of Hardy spaces \cite{Duren}, but we
leave this more mathematical matters for another work. 
In connection to the previous discussions we would
like to mention the work by Burnol, who
has emphasized the importance that causality in scattering theory
may play in the proof of the Riemann hpothesis \cite{Burnol1,Burnol2}.

\no Let us consider again some examples of potentials
inspired by the previous results.

\subsubsection*{Example 5: Other algebraic  potentials in the $\M_1$ model}

Let us choose $f_1(E)$ as

\beq
f_1(E) = C_1 + C_2 \prod_{n=1}^{2 M} \frac{ \alpha_n + i E}{ \alpha_n - i E}, 
\quad C_1 + C_2 = 1,  
\label{ej1}
\eeq

\no
where $\alpha_n > 0 \; (n=1, \dots, 2M)$ to avoid
the poles of $f_1(E)$ in the upper half plane.
The condition $C_{1} + C_2 = 1$
guarantees that $f_1(0) = 1$. From (\ref{j16}) one has

\beq
\hat{a}(E) = 
\frac{2 C_2}{i E} \left(  
\prod_{n=1}^{2 M} \frac{ \alpha_n + i E}{ \alpha_n - i E} -1 \right). 
\label{ej2}
\eeq

\no
The quantity in parenthesis is an analytic function
in $\Cmath_+$ which behaves as $1/E$ for $|E| >> 1$. 
The associated potential $a(q)$ can be computed from the Fourier transform
of $\hat{a}(E)$

\beq
a(q) = 4 C_2 \sum_{n=1}^{2 M} e^{- \alpha_n q} 
\; \prod_{m \neq n}^{2 M} 
\frac{ \alpha_m + \alpha_n}{ \alpha_m - \alpha_n}, 
\label{ej3}
\eeq

\no
and consists in the superposition of decaying exponentials. 
In terms of the $x$ variable ($\vep(x) = x$) the decay is algebraic. 
Using (\ref{ej1}) one can show that

\beq
\F_1(E) = 1 - 2 C_1  + 2 C_1 \; f_1(E),  
\label{ej4}
\eeq

\no
and in particular 

\beq
C_1 = C_2 = \frac{1}{2} \Longrightarrow 
\F_1(E) = f_1(E). 
\label{ej4b}
\eeq

\no 
The zeros of $\F_1(E)$ are given by

\beq
\prod_{n=1}^{2 M} \frac{ \alpha_n + i E}{ \alpha_n - i E}
= - \frac{1 - 2 C_1 + 2 C_1^2}{2 C_1 ( 1 - C_1)}.   
\label{ej5}
\eeq

\no
For $C_1= 1/2$ there are 
$2 M$  real solutions which appear in pairs $\{ E, - E \}$, 
while for other values
the solutions are complex and satisfy $\Im \; E < 0$ in agreement with 
(\ref{c65}). 
When  $M = 1$ and $C_1 = 1/2$, the potential (\ref{ej3}) becomes

\beq
a(q) = 2 \frac{ \alpha_2 + \alpha_1}{\alpha_2 - \alpha_1} 
\; ( e^{- \alpha_1 q} - e^{- \alpha_2 q}),   
\label{ej6}
\eeq

\no
and the solutions of (\ref{ej5}) are the pair
of energies

\beq
E = \pm \sqrt{ \alpha_1 \alpha_2}.  
\label{ej7}
\eeq

\subsubsection*{ Example 6: Other algebraic  potentials in the $\M_2$ model}

A general choice of the potentials $a$ and $b$ for the
$\M_2$ model is given by

\beq
a(q) = \sum_{n=1}^{N_a}  a_n e^{ - \alpha_n q}, \quad
b(q) = \sum_{m=1}^{N_b}  b_m e^{ - \beta_m q}, 
\label{ej8}
\eeq

\no
where $\alpha_n, \beta_m > 0$. The $S$-functions
can be easily computed using

\beq
S_{f_1, f_2} = - \frac{ \mu_1}{ 2( \mu_1 + \mu_2)} \; 
\frac{E}{ E + i \mu_1}, \quad f_i = e^{ - \mu_i q} \; \; (i=1,2).   
\label{ej9}
\eeq

\begin{figure}[t!]
\begin{center}
\includegraphics[height= 4 cm,angle= 0]{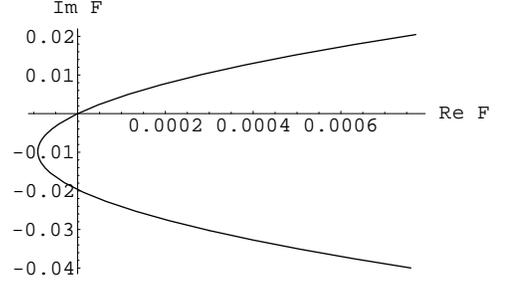}
\end{center}
\caption{
Plot of the real and imaginary parts of $\F(E)$ 
associated to the potentials (\ref{ej8}) 
and (\ref{ej10}). $E$ varies in the interval
$(1.91, 2.05)$. The curve passes through the origin
at $E = 1.95634$. 
}
\label{cruce3-24J}
\end{figure}

\no
Fig. \ref{cruce3-24J} displays an example with $N_a=2 $ and $N_b=1$
and the following values of the parameters in (\ref{ej8})

\barray 
& \alpha_1 = 1, \; \;  \alpha_2 = 4, \; \;  \beta_1 = 0.0360157, & 
\label{ej10} \\
& a_1 = 7.22928, \; \; a_2 = - 7.03245, \; \;  b_1 = 1. & 
\nonumber 
\earray 

\no
At $E=  \pm 1.95634$ the Jost function vanishes. This example
is a perturbation of the potential (\ref{ej3}) with
$\alpha_1 = 1$ and $\alpha_2 = 4$, which has bound states
at $E = \pm 2$. The most interesting feature of this example
is that  $\Re \; \F(E)$
becomes negative in a small neighbour of the origin. 
This has been possible by the  addition of the $b$ potential. 
Notice that $a_1, a_2$  do not balance
exactly as in eq.(\ref{ej6}).

\section{The Riemann zeta function and the Jost function}

There are two  main physical approaches to the Riemann zeros,
either as a bound state problem, or as a scattering problem.
In the former approach one looks for a Hamiltonian whose
point like spectrum is given by the Riemann zeros,
while in the latter the Riemann zeta function gives
the scattering amplitude of a physical system, whose
properties reflect in some way or another the existence of the zeros.  
Both approaches would  naturally converge if the
Riemann  zeros were the zeros of a Jost function
as suggested above. 

\begin{figure}[t!]
\begin{center}
\hbox{\includegraphics[height= 4.5 cm, width= 4.5 cm]{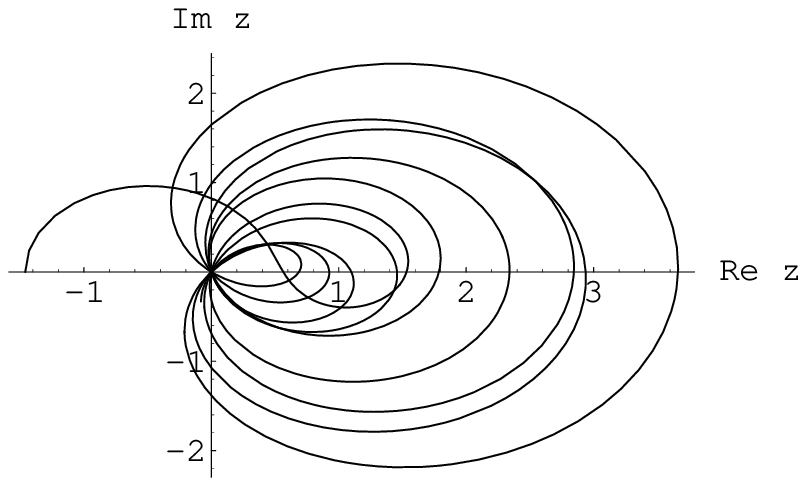}
\includegraphics[height = 4.5 cm, width= 3.5 cm]{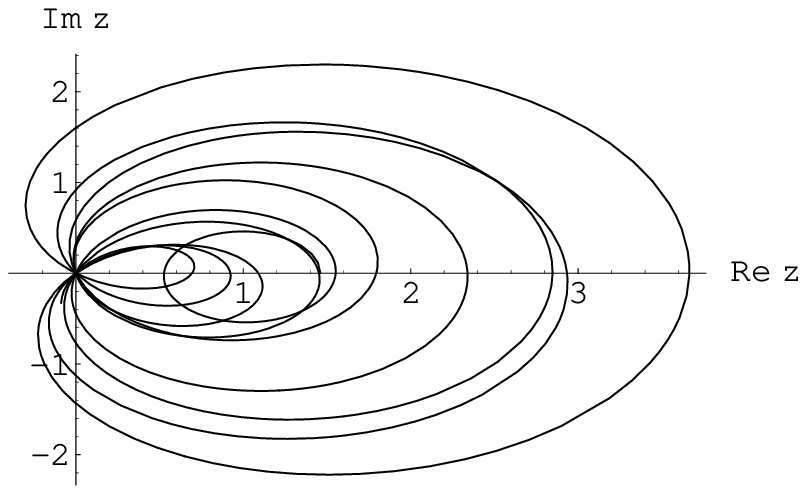}}
\end{center}
\caption{Left: real and imaginary parts of $\zeta(s)$
with $s= 1/2 - i E$ and $E \in (0, 50)$.
Right: same as before for  $\zeta_H(s) = (s-1) \zeta(s)/s$. 
}
\label{zeta1}
\end{figure}

The scattering approach was pionered by Faddeev and Pavlov
in 1975, and has been followed by many authors 
\cite{Faddeev,Lax,G2,Joffily}. An important result
is that the phase of $\zeta(1 + i t )$ is related
to the scattering phase shift of a particle moving 
on a surface with constant negative curvature. The chaotic nature of that
phase is a well known feature. Along this
line of thoughts, Bhaduri, Khare and Law (BKL) maded in 1994 
an analogy between  resonant quantum scattering amplitudes 
and the Argand diagram of the zeta function
$\zeta(1/2 - i t)$, where
the real part of $\zeta$ (along the $x$-axis) 
is plotted against the imaginary part ($y$-axis) \cite{BKL}. 
The diagram consists of an infinite series of closed loops
passing through the origin every time $\zeta(1/2 - i t)$
vanishes (see fig. \ref{zeta1}). This loop structure
is similar to the Argand plots of  partial wave amplitudes
of some physical models with the two axis being interchanged.
However the analogy is flawed since the real part of $\zeta(1/2  - it)$
is negative in small regions of $t$, a circumstance
which never occurs in those physical systems.

In fact, the loop structure of the models proposed by BKL
is identical, up to a scale factor of 2, to the model
of  example 1 (see fig. (\ref{step1})), where 
the loops representing $\F_1(E)$, for $a_1 = 1$,  
are circles of radius 1/2, centered
at $x=1/2$. For general models of type $\M_1$, the loops
are not circles but 
the real part of $\F_1(E)$ is always positive (see eq.(\ref{c63})), 
and therefore they can never represent $\zeta(1/2 - i E)$.
Incidentally, this constraint does not apply to the models
of type $\M_2$, where $\Re \; \F(E)$ may become negative,
as in the example 6 (see fig. \ref{cruce3-24J}). 
This suggests that
$\zeta(1/2 - i E)$ could be perhaps the Jost
function $\F(E)$ of a $\M_2$ model for a particular choice
of $a$ and $b$. However,  
$\zeta(1/2 - i E)$
has a pole in the upper half plane at $E = i/2$, 
while  $\F(E)$ is always analytic in that region.
The solution of this problem consists in moving the pole
to the lower half plane defining the function

\beq
\zeta_H(s) = \frac{s-1}{s} \zeta(s), 
\label{r1}
\eeq

\no
which has a pole at $E=- i/2$ for $s = 1/2 - i E$. 
This function was already considered by Hardy, 
and it is discussed in detail by 
Burnol in his approach to the RH \cite{Burnol1,Burnol2}. 
 Fig.\ref{zeta1} shows the Argand plot of 
$\zeta_H(1/2 - i t)$. In the rest of this section
we shall explore the possibility that the zeta
function, or some other related function, could be realized
as a Jost function.

\subsection{The Bessel potentials and the
 smooth part of the Riemann formula}

The zeta function satisfies the well know
functional equation,

\beq
\zeta(\frac{1}{2} - i t) =  \pi^{- i t} 
\frac{ \Gamma\left( \frac{1}{4} + \frac{i t}{2} \right) }
 { \Gamma\left( \frac{1}{4} - \frac{i t}{2} \right) } \; 
\zeta(\frac{1}{2} + i t). 
\label{r3b}
\eeq

\no 
For $t$ real, one defines

\beq
\zeta(1/2 + i t) = Z(t) \; e^{- i \theta(t)},  
\label{r2}
\eeq

\no
where $Z(t)$ is the Riemann-Siegel zeta function, which
is  even and real, and  
$\theta(t)$ is a phase angle given by

\beq
e^{2 i \theta(t)} = \pi^{- i t} 
\frac{ \Gamma\left( \frac{1}{4} + \frac{i t}{2} \right) }
 { \Gamma\left( \frac{1}{4} - \frac{i t}{2} \right) }, 
\label{r3}
\eeq

\no
which is taken to be continuous across  the Riemann zeros.
This angle gives   the smooth
part of the Riemann formula (\ref{s5}), i.e.

\beq
\langle \CN (t) \rangle = \frac{ \theta(t)}{\pi} + 1. 
\label{r4}
\eeq

\no 
As noticed by BKL, the loop structure depicted in fig. \ref{zeta1} 
shows that the zeros of $Z(t)$ are near to the
points where $\zeta(1/2 + it)$ is purely imaginary, i.e.
$\theta(t) = \pi (n + 1/2)$ \cite{BKL}. This observation suggests  
an approximation to the Riemann zeros

\beq
\cos \theta(t) = 0 \Longrightarrow 1 +  \pi^{- i t} 
\frac{ \Gamma\left( \frac{1}{4} + \frac{i t}{2} \right) }
 { \Gamma\left( \frac{1}{4} - \frac{i t}{2} \right) } = 0, 
\label{r5}
\eeq

\begin{figure}[t!]
\begin{center}
\includegraphics[height= 4.5 cm,angle= 0]{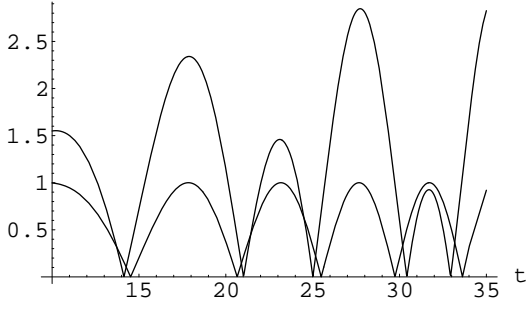}
\end{center}
\caption{
Comparison between the absolute values of 
$\zeta(1/2 + i t)$
and $\cos \theta(t)$. Observe the proximity of the points
where both quantities vanish.           
}
\label{zeta_abs}
\end{figure}

\no
which  works within a 3 $\%$  of error (see fig. \ref{zeta_abs})
and it is essentially the same as the smooth approximation
discussed in section II. 
Condition (\ref{r5}) was also obtained by Berry 
from the first term in his approximate formula
\cite{B-chaos}. 
BKL related $\theta(t)$ to the scattering phase
shift of a non relativistic particle moving in an inverted harmonic
oscillator, $H = p^2 - x^2$, 
a problem which is related to the
$H = x p$ by a canonical transformation.

In this section we shall  relate  $\theta(t)$ to a 
$\M_1$ model, with $\vep(x) = x$, and  potential

\beq
a(x) = c \; J_\nu( \lambda x), \quad 1 \leq x \leq \infty,  
\label{r6}
\eeq

\no
where $J_\nu$ is the Bessel function of order $\nu$,  and
$c$ and $\lambda$ are parameters to be fixed later on.
The Mellin transform of (\ref{r6}) (i.e. Fourier for $a(q)$)
yields,

\barray
&  \ha(t)  =   \int_1^\infty dx \; c \;  x^{- 1 + i t} \;  J_\nu( \lambda x) 
& \label{r7} \\
&  =  c \;  2^{-1 + i t} \lambda^{- i t}
\; \frac{ \Gamma \left( \frac{\nu + i t}{2} \right) } 
{ \Gamma \left( 1+ \frac{\nu - i t}{2} \right) } \nonumber & \\
& 
- \frac{ c \; 2^{ - \nu} \lambda^{ \nu} }{ ( \nu + i t) \Gamma( 1 + \nu)}
\; _1F_2 \left( \frac{ \nu + i t}{2}; 1 +  \frac{ \nu + i t}{2}, 1 + \nu; -
\frac{ \lambda^2}{4}  \right), &  
\nonumber 
\earray

\no
where $_1F_2$ is a hypergeometric function of type (1,2) \cite{GR}. 
By the Titchmarsh theorem,  $\ha(t)$ is an analytic function in the upper
half plane. Indeed, the poles of the gamma function 
in the numerator of the first term are cancelled out by the poles
of the second term. 
In the limit where $|t| \rightarrow \infty$, one gets

\barray 
& & \lim_{|t| \rightarrow \infty} \; 
 _1F_2 \left( \frac{ \nu + i t}{2}; 1 +  \frac{ \nu + i t}{2}, 1 + \nu; -
\frac{ \lambda^2}{4}  \right)
\nonumber  \\
& &  = 
\; _0F_1 \left( 1 + \nu,- \frac{ \lambda^2}{4}  \right), 
\label{r8}
\earray

\no
which is related to the Bessel function

\beq
J_\nu(z) = \frac{ (z/2)^\nu}{ \Gamma(1 + \nu)}  \; 
 _0F_1 \left( 1 + \nu,- \frac{ z^2}{4}  \right),  
\label{r9}
\eeq

\no
thus

\beq
\ha(t) \sim \frac{i c}{t} 
\left[ (\lambda/2)^{- i t}  
\frac{ \Gamma \left( \frac{\nu + i t}{2} \right) } 
{ \Gamma \left(\frac{\nu - i t}{2} \right) } 
+ J_\nu(\lambda) \right], \quad  |t| >> 1, 
\label{r10}
\eeq

\no
and consequently

\beq
f_1(t) = 1 + \frac{i t}{2} \,  \ha(t) \sim 
1 -  \frac{c}{2} 
\left[ (\lambda/2)^{- i t}  
\frac{ \Gamma \left( \frac{\nu + i t}{2} \right) } 
{ \Gamma \left(\frac{\nu - i t}{2} \right) } 
+ J_\nu(\lambda) \right].
\label{r11}
\eeq

\no 
A necessary condition for $\F_1(t)$ to vanish 
is that $f_1(t)$ vanish as well, which 
in the limit $|t| >> 1$ is guaranteed by

\beq
f_1(t) \sim 0 \Longrightarrow 
\frac{2}{c} -  J_\nu(\lambda) = 
 (\lambda/2)^{- i t}  
\frac{ \Gamma \left( \frac{\nu + i t}{2} \right) } 
{ \Gamma \left(\frac{\nu - i t}{2} \right) }  = \pm 1.  
\label{r12}
\eeq

\no
This equation coincides with  
(\ref{r5}), if we take the minus sign in the RHS 
in (\ref{r12}) and choose

\beq
\nu = \frac{1}{2}, \quad \lambda = 2 \pi, \quad c = -2.    
\label{r13}
\eeq

\no
Since $J_{1/2}(x) = \sqrt{2/ \pi x} \sin(x)$, 
the corresponding potential is

\beq
a(x) = - \frac{2}{\pi} \; \frac{ \sin( 2 \pi x)}{  \sqrt{x}}. 
\label{r14}
\eeq

\begin{figure}[t!]
\begin{center}
\includegraphics[height= 4 cm,angle= 0]{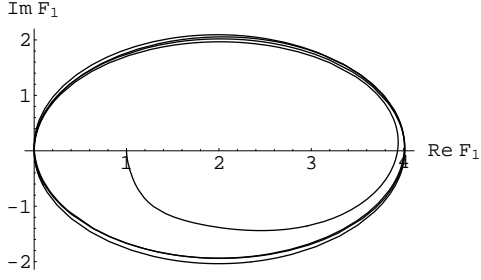}
\end{center}
\caption{
Plot of the numerical evaluation of 
$\F_1(t)$ for the potential (\ref{r14})
in the range $0 < t < 30$. 
}
\label{F1-bessel}
\end{figure}

\no
The Jost function $\F_1(t)$ is found numerically 
using equation eq.(\ref{j17}), which involves 
the Hilbert transform of $|f_1(t)|^2$. Fig. \ref{F1-bessel}
shows the Argand plot of $\F_1(t)$,
which consists of a series of loops passing 
very close to the origin, at those values
of $t$  accurately approximated
by eq.(\ref{r5}). For $|t| >> 1$ the loops are
circles of radius 2 centered at $x = 2$.
This numerical result can be obtained analytically. 
After a long computation one finds that

\beq
\F_1(t) = 1 - c \;  e^{2 i \theta(t)}
  + \frac{c^2}{4} + O(\frac{1}{t}),   
\label{r14d}
\eeq

\no
which for $c = -2$ becomes,

\beq
\F_1(t) = 4 \cos \theta(t) \; e^{ i \theta(t)} +  O(\frac{1}{t}). 
\label{r14e}
\eeq

\no 
It is interesting to compare (\ref{r14e})
with (\ref{r2}), which we write as

\beq
\zeta(1/2 - i t)  = Z(t) \; e^{i \theta(t)}.  
\label{r14e-b}
\eeq

\no
Up to $1/t$ terms, the phase factor is the same,
while $Z(t)$ is replaced by $\cos \theta(t)$,
which is precisely the approximation that reproduces
the smooth part of the zeros (\ref{r5}). Hence, to leading order
in $1/t$, the Jost function (\ref{r14e})
can be considered as the smooth approximation
to  $\zeta(1/2 - i t)$.

\subsection{Relation to the Berry-Keating regularization}

The Bessel function $J_\nu(x)$ satisfies the second order diferential
equation

\beq
\left( x^2 \frac{d^2}{dx^2} + x  \frac{d}{dx} + x^2 - \nu^2 
\right) \; J_\nu (x)  = 0, 
\label{r15}
\eeq

\no
which can be rewritten as

\beq
( x \; p)^2  J_\nu (\lambda x) = \hbar^2 ( \lambda^2 \;  x^2 - \nu^2)
  J_\nu (\lambda x),
\label{r16}
\eeq

\no
where $p = - i \hbar \frac{d}{dx}$. From the 
definition (\ref{c1-4}),

\beq
\psi_a(x) = \frac{c J_\nu(\lambda x)}{\sqrt{x}} = 
\frac{c \, \sin( 2 \pi x)}{\pi x}, 
\label{r17}
\eeq

\no
eq. (\ref{r16}) turns into

\beq
H_0^2  \;  \psi_a (x) = \hbar^2 ( \lambda^2 \;  x^2 - \nu^2)
  \psi_a(x), 
\label{r18}
\eeq

\no
where  $H_0 = \sqrt{x} p \sqrt{x}$ is the BKC Hamiltonian
(\ref{qu2}).  Dropping  $\psi_a$ in both sides
and replacing $H_0$ by $ xp$, one obtains a 
 classical version of  (\ref{r18}),

\beq
(x p )^2 =  \hbar^2( \lambda^2 \;  x^2 - \nu^2) \Longrightarrow
p = \pm \hbar \lambda \sqrt{ 1 - \frac{\nu^2}{(\lambda x)^2}},  
\label{r19}
\eeq

\begin{figure}[t!]
\begin{center}
\includegraphics[height= 4 cm,angle= 0]{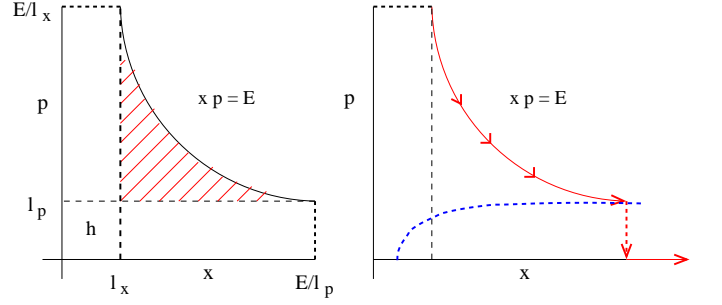}
\end{center}
\caption{
Left: allowed phase space region in the BK semiclassical
quantization. Right: the dotted line represents the
classical version of the state $\psi_a$  (\ref{r19}). 
}
\label{my_interpret}
\end{figure}

\no
which describes a curve in  phase space that approaches asymptotically
the lines $p = \pm \hbar \lambda$. 
We shall identify
these  asymptotes with the BK boundary 
in the allowed momenta $|p| = l_p$.
(see fig. \ref{my_interpret}). 
Recall on the other hand  the
boundary condition $ x \geq l_x = 1$, which 
combined with the previous identification 
reproduces the Planck cell quantization condition,

\beq
l_p = \hbar \lambda, \;\; l_x = 1
\Longrightarrow l_p \,  l_x = \hbar \lambda = 2 \pi \hbar 
\label{r20}
\eeq

\no
where we used that $\lambda = 2 \pi$. 
This interpretation of the state
  $\psi_a(x)$ shows that the BK boundary 
$|p| = l_p$ is realized in our model
in a dynamical way and not as a constraint
in phase space.

To complete this picture, let us try to 
 understand the physical meaning of the state $\psi_b$ for $b(x) = 1$. 
Writing

\beq
\psi_b(x) = \frac{1}{ \sqrt{x} } \;  \sign(x-1),  
\label{r21}
\eeq

\no
one finds that its time evolution under $H_0$ is given by

\beq
\psi_b(x,t) = e^{-i t H_0} \psi_b(x)= \frac{1}{ \sqrt{x} }\; \sign(x-e^t),  
\label{r21-b}
\eeq

\no 
which is a kink state associated to the classical
trayectory

\beq
x(t) = x_0 \, e^t, \;\; p(t) = 0,   \;\; E = x(t) p(t) = 0
\label{r24}
\eeq

\no
 $H_0$ does not have a zero energy
eigenstate, but based on these results one can think
of $\psi_b$ heuristically as that state.
The $p =0$ line in phase space is the classical
analogue of the state $\psi_b$, just like the lines
$p = \pm l_p$ are the classical analogue of  $\psi_a$. 
This interpretation allows us to understand heuristically, 
 the BK-regularization. 
Indeed, take a particle with energy $E$, 
which at the initial time $t=0$ is at $x=1$ and $p = E$. 
Following the classical trayectories
(\ref{s2}) this particle will reach at time $t_1$, 
the boundary $p(t_1)= l_p$ and then suddenly loose
all its momenta,   $p(t_1+ \epsilon)= 0$
(see fig. \ref{my_interpret}).  
The phase
space area involved in this evolution agrees basically 
with the calculation
made by Berry and Keating. If this area is 
a integer multiple  
of $2 \pi \hbar$, then there is a bound state. 
In the previous argument, one should strictely reverse
the time arrow since the classical 
trayectories generated by $1/( x p)$ are the time-reversed
of eqs. (\ref{s2}), but the result does not change. 
At the quantum level 
the existence of bound states is due to an interference effect.
In the absence of this interference the boundary at $p = \pm l_p$
behaves as a  ``transparent'' wall,  and the particles do
not return to their initial position. This situation
corresponds to Connes picture where
all the eigenstates are delocalized. 
In this manner, the Berry-Keating and Connes pictures 
may coexist  in a coherent picture both semiclassically
and quantally.

Finally, we would like to make a comment concerning the
wave function (\ref{r17}), and its relation to 
the von Neumann and Wigner potentials, mentioned
in section IV \cite{Galindo,neumann-wigner,simon,arai,cruz}. 
A common feature of these potentials is their
asymptotic behaviour,  $\sin(r)/r$,
where $r >> 1$ is the radius. The potentials  
having a positive energy eigenstate
form a submanifold. So that 
a fine-tuning of  couplings
is required. It is interesting to observe
the similarity with the potential  (\ref{r17}),
and the sensitivity to the choice
of couplings in order to have bound states.

\subsection{Potentials for the Riemann zeros}

In reference \cite{BK1} Berry and Keating tried
to replace the semiclassical regularization of $ xp$
with quantum boundary conditions that would generate
a discrete spectrum. A  proposal is  to use the dilation
symmetry of $x p$, i.e.

\beq
x \rightarrow K \; x, \qquad 
p \rightarrow p/K,  
\label{r25} 
\eeq

\no
where $K$ corresponds to an evolution after time 
$\log K$ as indicated by eq.(\ref{s2}) 
(see  \cite{Aneva} for a discussion of the symmetries
of $H = xp$). 
The Hamiltonian $H_0$  (\ref{qu1}), is the 
generator of the scale transformations:

\beq
\psi(K \, x) = \frac{1}{K^{\frac{1}{2} - i H_0}} \;  \psi(x).   
\label{r26} 
\eeq

\no 
Based on this, BK considered a linear
superposition of the  wave function
$\psi_E(x)= C x^{- 1/2 + i E}$ with  the ones obtained
by integer dilations $K = m$,

\barray 
\psi_E( x) &    \rightarrow & \sum_{m=1}^\infty \psi_E(m \, x)
\label{r27} \\
& 
= &  \frac{C}{x^{1/2 - i E}} \; \sum_{m=1}^\infty \frac{1}{m^{1/2 - i E}}
=  \frac{C}{x^{1/2 - i E}} \; \zeta(\frac{1}{2} - i E),  
\nonumber 
\earray

\no
so that the vanishing of (\ref{r27}) could be interpreted
as an eigencondition. However, there is no justification
for that condition, nor it is clear its physical
or geometrical meaning. The approach we have been
following so far is to implement the boundary
conditions in a dynamical manner, hence
it is more natural to impose
the  symmetry under discrete dilations,
not on the eigenfunctions but on 
the potentials. Consider the linear superposition of
(\ref{r17}),

\barray 
\psi_a( x) &  \rightarrow &  \sum_{m=1}^\infty \psi_a(m \, x)
\label{r28} \\
& = &   \frac{c}{x} \sum_{m=1}^\infty \frac{  \sin( 2 \pi m x)}{\pi m}
=  \frac{c}{x} \left( [x] - x + \frac{1}{2} \right),  
\nonumber 
\earray

\no
where $[x]$ denotes the integer part of $x$. 
We have used the Fourier
 decomposition of the sawtooth function
$[x] - x + 1/2$. 
The Mellin transform
of the potential $a(x) = \sqrt{x} \psi_a(x)$
associated to (\ref{r28}) is  given by

\barray 
\ha(t) & = &  \int_1^\infty dx \, x^{-1 + i t} \;  a(x) 
\label{r29} \\
& = & 
\frac{c}{ \frac{1}{2} - i t} \left( 
\zeta(\frac{1}{2} - i t) + \frac{1}{\frac{1}{2} + i t} - 
\frac{1}{2} \right),  
\nonumber
\earray

\no
where we have used \cite{Edwards}

\beq
\zeta(s) = s 
\int_1^\infty dx \; \frac{  [x] - x + \frac{1}{2}}{ x^{s +1}}
+ \frac{1}{s-1} + \frac{1}{2}, \; \;  \Re \;  s > 0.   
\label{r30}
\eeq

\no
Choosing $b(x) =1$, the  Jost function $\F_1(t)$, 
associated to (\ref{r28}),  in leading order in $t$,
is given by

\begin{figure}[t!]
\begin{center}
\includegraphics[height= 6 cm,angle= 0]{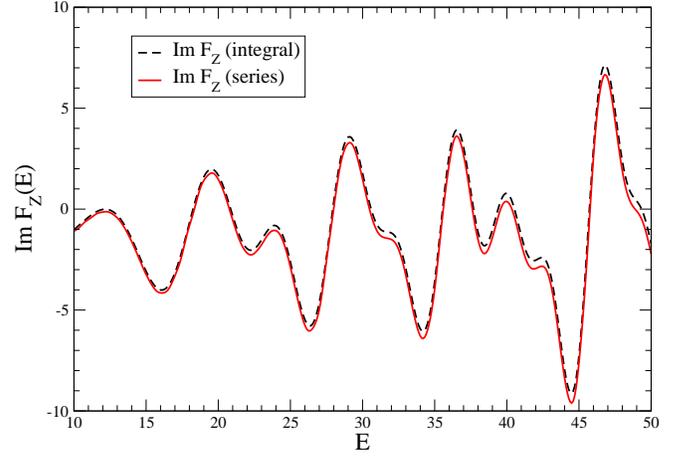}
\end{center}
\caption{
Numerical values of ${\rm Im} \;  \F_Z(E)$ using two methods:
1) integrating (\ref{r35}) in the interval $(0,d)$ with 
$d = 400$, and 2) Summing the series
(\ref{r36}) up to  $M=5000$.
}
\label{ImF-10-50}
\end{figure}

\beq
\F_1(t) \sim \left(1 + \frac{c}{4} \right)^2 -
c  \left(1 + \frac{c}{4}\right) \; \zeta(1/2 - i t)
- c^2 \; S_{a_0, a_0}(t),   
\label{r31}
\eeq

\no
where
$S_{a_0, a_0}$ denotes the function
 (\ref{j10}) with $\hat{f} = \hat{g}= a_0$, and

\beq
a_0(t) = \frac{\zeta(1/2 - i t)}{\frac{1}{2} - i t}.  
\label{r32}
\eeq

\no
If $c=-4$, the Jost function $\F_1$ is 
given asymptotically by $S_{a_0, a_0}$,

\beq
\F_1(E)/4 \sim \frac{ E^2 \; Z(E)^2}{ \frac{1}{4} + E^2}
- i  P \int_{- \infty}^\infty \frac{dt}{\pi} \;
\frac{ 1}{t - E} \; \frac{ E t \; Z(t)^2}{ \frac{1}{4} + t^2}.  
\label{r33}
\eeq

\no
A further approximation of  (\ref{r33}) is

\beq
\F_1(E)/4 \sim \F_{Z}(E) =  \,  Z(E)^2 
- i  P \int_{- \infty}^\infty \frac{dt}{\pi} \;
\frac{ \,  Z(t)^2}{t - E}, 
\label{r34}
\eeq

\no
or using that $Z(t) = Z(-t)$

\beq
\F_Z(E) =   \,  Z(E)^2 
- 2 i E  \;  P\int_{0}^\infty \frac{dt}{\pi} \;
\frac{ \,  Z(t)^2}{t^2 - E^2},  
\label{r35}
\eeq

\no
The Cauchy integral, giving the imaginary
part of $\F_Z(E)$, is convergent   
 thanks to the asymptotic behaviour  
$|Z(t)| \sim |t|^{1/4 + \epsilon} \;(\epsilon > 0)$, 
on the critical line \cite{Titchmarsh2}.
The integral (\ref{r35}) can be computed numerically 
in the interval $(0,d)$, with $d$ sufficiently large. 
The results converge rather slowly with $d$. 
Fig. \ref{ImF-10-50} shows ${\rm Im} \;  \F_Z(E)$  
for $d=400$ in the interval $E \in (10,50)$.

\no
An alternative method to find ${\rm Im} \;  \F_Z(E)$ 
is to Hilbert transform  $Z(t)^2$, 
using the well known series expansion of 
the zeta function,

\beq
\zeta(s) = \frac{1}{2^{1 - s} -1} \; \sum_{n=1}^\infty 
\frac{ (-1)^n}{ n^s}, \qquad {\rm Re} \; s > 0.  
\label{r36}
\eeq

\no
After a long calculation one finds,

\barray
& {\rm Im} \, \F_Z(t) =  -i \lim_{M \rightarrow \infty}   & \label{r37} \\
& 
\left[ 
\frac{1}{p(t) \; p(-t)} \; \sum_{n,m}^M
\frac{(-1)^{n+m} \; \sign(n-m) }{ n^{1/2 - i t} \; m^{1/2 + i t} } 
\right.  & \nonumber \\
& +   \frac{p(t) - p(-t)}{p(t) p(-t)} \sum_{n=1}^M \frac{1}{n} & 
\nonumber  \\
&   \left. - \sum_{n > m=1}^M \frac{ 2 \; (-1)^{n+m} }{n}
\left( \frac{ 2^{( 1/2 + i t) \{ \log_2(n/m) \} }}{p(t)}
- ( t \rightarrow -t) 
\right) 
\right],  &  
\nonumber 
\earray

\no
where $p(t) = 2^{1/2 + i t} -1$ and $ \{ \log_2(n/m) \}$
is the fractional part of  $\log_2(n/m)$.
Fig.\ref{ImF-10-50}
shows the values of  ${\rm Im} \; \F_Z(E)$ 
computed with eq.(\ref{r37}), for $M = 5000$
in the interval $E \in (10,50)$, which agrees
reasonable well with the result obtained with the 
truncated integral (\ref{r35}). 
For larger values of $E$ it is more convenient to use
the series expansion (\ref{r37}). 
The complete expression of  $\F_Z(t)$ is obtained   adding to 
(\ref{r37}), its real part  $Z(t)^2$,

\begin{figure}[t!]
\begin{center}
\includegraphics[height= 6 cm,angle= 0]{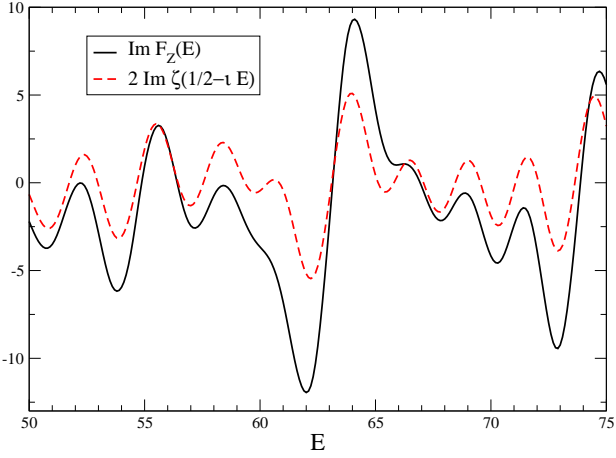}
\end{center}
\caption{ 
Numerical values of ${\rm Im} \;  \F_Z(E)$ and
${\rm Im} \; \zeta(1/2 - i t)$ in the interval
$E \in (50,75)$. 
}
\label{ImF-50-75}
\end{figure}

\barray
&  \F_Z(t) =   \lim_{M \rightarrow \infty}   & \label{r38} \\
& 
\left[ 
\frac{2}{p(t) \; p(-t)} \; \sum_{n > m}^M
\frac{(-1)^{n+m}}{ n^{1/2 - i t} \; m^{1/2 + i t} } 
\right.  & \nonumber \\
& +   \frac{1 + p(t) - p(-t)}{p(t) p(-t)} \sum_{n=1}^M \frac{1}{n} & 
\nonumber  \\
&   \left. - \sum_{n > m=1}^M \frac{ 2 \; (-1)^{n+m} }{n}
\left( \frac{ 2^{( 1/2 + i t) \{ \log_2(n/m) \} }}{p(t)}
- ( t \rightarrow -t) 
\right) 
\right],  &  
\nonumber 
\earray

As expected, eq. (\ref{r38})
does not have poles in the upper half plane.
The poles arising from the terms proportional to
 $1/p(t)$ cancell each other.

\begin{figure}[t!]
\begin{center}
\includegraphics[height= 6 cm,angle= 0]{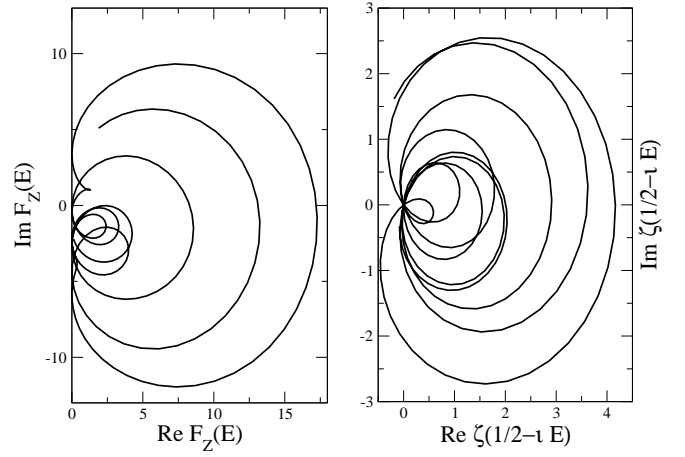}
\end{center}
\caption{
Argand plane representation of  $\F_Z(E)$ (left) 
and $\zeta(1/2 - i E)$ (right) in the region
$ E \in (50,75)$. 
The loops associated to  $\F_Z(E)$
do not generally passed through the origin and
the are slightly displaced downward. 
}
\label{loops-50-75}
\end{figure}

The real part of $\F_Z(E)$ vanishes at the zeros
of $Z(E)$. The question is wether
its imaginary part, given by (\ref{r37}), also does.
The answer to this question is negative in general, as
can be seen  from 
fig. \ref{ImF-50-75}, which  plots the values
of   ${\rm Im}\; \F_Z(E)$,  and those of
$2 \;  {\rm Im} \zeta(1/2 - i E)$ in the interval $(50,75)$.
Observe that the shape of the two curves
is similar, but their zeros do not coincide.
Curiously enough, their maxima and minima are 
much closer.  Fig. \ref{loops-50-75}
displays the Argand plot of  $\F_Z(E)$
and $\zeta(1/2 - i E)$ in the interval $E \in (50,75)$.
Observe again the similarity between their loop
structures.  We know of  no reason why
  ${\rm Im}\; \F_Z(E)$ should vanish, even
asymptotically, at the zeros of 
$ {\rm Im} \;  \zeta(1/2 - i E)$ or $Z(t)$.
If that were the case, then the Riemann zeros
would become resonances with a life-time
increasing asymptotically, but this seems unlikely. 
Our conclusion is that $\psi_a(x)$  (\ref{r28}) 
is not enough
to yield the Riemann zeros in the  spectra, 
and that one needs a non trivial potential $b(x)$.
A concrete proposal is to look for a $\M_2$
model yielding a Jost function $\F(t)$ proportional
to $\zeta_H(1/2 - i t)$ ( recall (\ref{r1})). 
We do not see at the moment any obstruction for
this realization, but one needs additional
insights.

We shall end this section with some comments
and suggestions.

\begin{itemize}

\item The function $\F_Z(t)$ (\ref{r38}), 
reminds a two-variable version of the zeta
function first proposed by  Euler

\beq
\zeta(s_1, s_2) = 
 \sum_{n_1 > n_2 > 0}^\infty 
\frac{1}{ n^{s_1}_1 \; n^{s_2}_2}, 
\label{r39}
\eeq

\no
if one chooses $s_1 = 1/2 - i t$ and 
$s_2 = 1/2 + i t$. Eq.(\ref{r39}),
together with its multivariable
extension, called Euler-Zagier zeta functions,
have attracted much attention 
in various fields, as  knot theory,
perturbative quantum field theory, etc
(see \cite{euler-zagier-1},  \cite{euler-zagier-2}
and references therein). 
The function $\zeta(s_1, s_2)$  satisfies
the so called shuffle relation

\beq
\zeta(s_1, s_2) + \zeta(s_2, s_1)
=  \zeta(s_1)  \zeta(s_2) -   \zeta(s_1+ s_2),
\label{r40}
\eeq

\no
which amounts in our case to the 
condition $\F_Z(t) + \F_Z(-t) = 2 Z(t)^2$. 
The two variable Euler-Zagier function 
can be meromorphically continued to $\Cmath^2$,
except at the singularities $s_1 = 1$ and 
$s_1 + s_2 = 2,1,0, -2, -4, \dots$. 
Hence, the identification $s_{1,2} = 1/2 \mp i t$
is  singular and a proper definition of 
$\zeta(s_1, s_2)$  requires a renormalization, 
probably along the lines of reference \cite{euler-zagier-2}
using Hopf algebras.

\item  The results obtained in this section 
can be generalized
to Dirichlet $L$-functions with real characters $\chi$ 
\cite{Davenport}. 
We summarize briefly the main results. From the functional
relation satisfied by the $L$-functions, it follows that 
the potentials, reproducing asymptotically the smooth
positions of the zeros of $L(s,\chi)$, are given by

\beq
a(x) \propto
\left\{ 
\begin{array}{ccc}
\sin (\lambda x)/\sqrt{x}, &  & \chi: \; {\rm even} \\
\cos (\lambda x)/\sqrt{x}, &  & \chi: \; {\rm odd} \\
\end{array}
\right. , 
\label{r41}
\eeq

\no
which corresponds to the Bessel functions 
$J_\nu(\lambda x)$ with $\nu = \pm 1/2$. 
The value $\lambda$ is related to the period of the
character $f$ as

\beq
\lambda = \frac{2 \pi}{f}, \qquad \chi(n + f) = \chi(n), \;\;
\forall n \in \Nmath. 
\label{r42}
\eeq

\no
The generalization of (\ref{r28}) is

\beq
\psi_a( x) \rightarrow \sum_{m=1}^\infty \chi(m) 
\psi_a(m \, x) = \frac{c}{x} 
\sum_{m=1}^\infty 
\frac{  \chi(m) }{m}
\left\{ 
\begin{array}{c}
\sin (\lambda m x)   \\
\cos (\lambda m x)    \\
\end{array}
\right. , 
\label{r43} 
\eeq

\no
so that the characters $\chi(m)$ must be real. 
The corresponding Mellin transform of $a(x) = x^{1/2}
\psi_a(x)$ is proportional to the Dirichlet function
$L(1/2 - i t, \chi)$. As in the case of the zeta function, 
one also needs non trivial $b$ potentials 
to relate the Dirichlet functions to the Jost functions
of the $\M_2$ model.

\end{itemize}

\section{Conclusions}

In this paper we have proposed
a possible realization of the Hilbert-P\'olya
conjecture in terms of a Hamiltonian given by a perturbation
of $H = xp$, or rather its inverse $1/(x p)$, 
 by means of an antisymmetric matrix
parameterize by two potentials $a(x)$ and $b(x)$.
The Schr\"odinger equation can be reduced
to a first order differential equation, suplemented
with boundary conditions, which are exactly solvable 
in terms of a Jost function.
In this respect, our approach is essentially different
from the second order approaches to the RH based
on standard QM.

The generic spectrum 
consists in a continuum of eigenenergies
which may contain a point like spectra embbeded in it. 
We have studied a variety of examples showing 
that the existence of a point like spectrum depends
``critically'' on the values of the coupling constants
of the model. We have found the potentials whose
resonances approach the smooth Riemann zeros 
asymptotically. In the classical limit 
these potentials reproduce the Berry-Keating 
semiclassical regularization 
of $H = x p$. Implementing 
a discrete dilation symmetry on the previous
potentials, we have obtained a Jost function
which, in the asymptotic limit, 
resembles the two-variable 
Euler-Zagier zeta function 
$\zeta(s_1, s_2)$ with $s_{1,2} = 1/2 \mp i t$. 
The real part of this Jost function vanishes
at the Riemann zeros but not necessarily
its imaginary
part. These results were derived from a trivial
potential $b(x)=1$, and they suggest
that a non trivial choice of $b(x)$,
could yield a Jost function
directly related to the zeta function.
A natural candidate is 
$ \frac{s-1}{s} \zeta(s)$, with $s= 1/2 - i t$,
which has the correct analiticity properties.
If these potentials do exist, then 
the Riemann zeros would become bound states 
of the model and the 
RH would follow automatically. 
This is how the  Hilbert-P\'olya conjecture 
would come true in our approach.

\section*{Acknowledgments}
I wish to thank Andre LeClair 
for the many discussions we had on our joint work on the 
Russian doll Renormalization Group 
and its relation to the Riemann hypothesis. 
I also thank M. Asorey, L.J. Boya,  J. Garc\'{\i}a-Esteve, 
M.A. Mart\'{\i}n-Delgado, G. Mussardo, 
J. Rodr\'{\i}guez-Laguna and J. Links for our conversations.  
This work was supported by the 
CICYT of Spain
under the contracts BFM2003-05316-C02-01
and FIS2004-04885.  I also acknowledge 
ESF Science Programme INSTANS 2005-2010.


\begin{thebibliography}{999}





\bibitem{Edwards} H.M. Edwards, ``Riemann's Zeta Function'', 
Academic Press, New York, 1974. 


\bibitem{Titchmarsh2} E.C. Titchmarsh, ``The Theory of the Riemann 
Zeta-Function'', 2nd ed., Oxford University Press 1999, Oxford. 


\bibitem{Bombieri} E. Bombieri, ``Problems of the Millenium: the Riemann
hypothesis'', Clay Mathematics Institute (2000). 


\bibitem{Sarnak} P. Sarnak, ``Problems of the Millenium: the Riemann
hypothesis (2004)'', Clay Mathematics Institute (2004). 

\bibitem{Conrey} J.B. Conrey, 
"The Riemann Hypothesis." Not. Amer. Math. Soc. 50, 341-353, 2003.


\bibitem{Watkins} See M. Watkins at 
http://secamlocal.ex.ac.uk/$\sim$mwatkins

/zeta/physics.htm
for a comprehensive review on several approaches to the RH. 


\bibitem{Rosu}  H.C. Rosu, ``Quantum hamiltonians and prime numbers'', 
Mod. Phys. Lett. {\bf A18} (2003) 1205; quant-ph/0304139. 

\bibitem{Elizalde} E. Elizalde, V. Moretti, S. Zerbini, 
 ``On recent strategies proposed for proving the Riemann hypothesis'', 
Int.J.Mod.Phys. {\bf A18} (2003) 2189-2196; math-ph/0109006.



\bibitem{Selberg} A. Selberg, 
"Harmonic analysis and discontinuous groups in 
weakly symmetric Riemannian spaces with applications 
to Dirichlet series", Journal of the Indian Mathematical 
Society 20 (1956) 47-87. 


\bibitem{Mont} H. Montgomery, ``The pair correlation 
of zeros of the zeta function'', Analytic Number Theory,
AMS (1973). 


\bibitem{Odl} A. Odlyzko, ``On the distribution of spacings
between zeros of zeta functions'', Math. Comp. {\bf 48}, 273 (1987).  


\bibitem{B-chaos} M.V. Berry, in {\em Quantum Chaos and Statistical
Nuclear Physics}. Eds. T.H. Seligman and H. Nishioka, Lecture 
Notes in Physics, No. 263, Springer Verlag, New York, 1986. 


\bibitem{Gutzwiller} M. C. Gutzwiller 
"Periodic orbits and classical quantization conditions",
J. Math. Phys. 12 no. 3 (1971). 



\bibitem{Julia} B. Julia, ``Statistical theory of numbers'',
in Number Theory and Physics, Springer Proceedings in Physics,
{\bf 47} (1990). 

\bibitem{BC} J.-B. Bost and A. Connes, ``Hecke algebras,
Type III factors and phase transitions with spontaneous
symmetry breaking in number theory''. Selecta Mathematica, 
New Series {\bf 1}, No. 3, 411, (1995). 


\bibitem{Mussardo} G. Mussardo, 
``The Quantum Mechanical Potential for the Prime Numbers'',
cond-mat/9712010. 





\bibitem{BK1} M.V. Berry and J.P. Keating, 
``H=xp and the Riemann zeros'', 
in {\em Supersymmetry and Trace Formulae: Chaos and Disorder}, 
ed. J.P. Keating, D.E. Khmelnitskii and 
I. V. Lerner, Kluwer 1999. 


\bibitem{BK2} M. V. Berry and J. P.  Keating, 
``The Riemann zeros and eigenvalue asymptotics'',  
SIAM REVIEW {\bf 41} (2) 236, 1999.  


\bibitem{Connes} A. Connes,  ``Trace formula in noncommutative 
geometry and the zeros of the Riemann zeta function'', 
 Selecta Mathematica (New Series) 5 (1999) 29; 
math.NT/9811068.

\bibitem{JSTAT} G. Sierra, ``The Riemann zeros and the Cyclic
Renormalization Group'', J.Stat.Mech. 0512 (2005) P006;
math.NT/0510572. 


\bibitem{RD1} A. LeClair, J.M. Rom\'an and G. Sierra, 
``Russian doll Renormalization Group and Superconductivity'', 
Phys. Rev. {\bf B69} (2004) 20505; cond-mat/0211338. 


\bibitem{RD2} A. Anfossi, A. LeClair, G. Sierra, 
``The elementary excitations of the exactly solvable 
Russian doll BCS model of superconductivity'', 
J. Stat. Mech. (2005) P05011; cond-mat/0503014. 


\bibitem{links} C.  Dunning and J. Links,
``Integrability of the Russian doll BCS model'',
Nucl. Phys. {\bf B702} (2004) 481, cond-mat/0406234.






\bibitem{GW} S. D. Glazek and K. G. Wilson, ``Limit cycles in
quantum theories'', Phys. Rev. Lett. {\bf 89} (2002) 230401,
hep-th/0203088;
``Universality, marginal operators, and limit cycles'',
Phys. Rev. {\bf B69}, 094304 (2004); 
cond-mat/0303297.


\bibitem{BLflow}  D. Bernard and  A. LeClair, 
``Strong-weak coupling duality in anisotropic current interactions'', 
Phys.Lett. {\bf B512} (2001) 78; hep-th/0103096. 

\bibitem{nuclear} P. F. Bedaque, H.-W. Hammer, and U. van Kolck,
``Renormalization of the Three-Body System with Short-Range
Interactions'', Phys. Rev. Lett. {\bf 82} (1999) 463, nucl-th/9809025.


\bibitem{fewbody}  E.  Braaten and  H.-W. Hammer,
`` Universality in Few-body Systems with Large Scattering Length'',
Phys.Rept. 428 (2006) 259-390, 
cond-mat/0410417.


\bibitem{morozov} A. Morozov and A. J. Niemi,
``Can Renormalization Group Flow End in a Big Mess?'', 
Nucl. Phys. {\bf B666}, 311 (2003); 
hep-th/0304178.



\bibitem{LRS1} A. LeClair, J.M. Rom\'an and G. Sierra,
``Russian doll Renormalization Group 
and Kosterlitz-Thouless Flows'', 
Nucl. Phys. {\bf B675}  (2003) 584; hep-th/0301042. 



\bibitem{LRS2} A. LeClair, J.M. Rom\'an and G. Sierra,
``Log-periodic behaviour of finite size effects in 
field theory models with cyclic renormalization group'', 
Nucl. Phys. {\bf B700} [FS] (2004) 407; hep-th/0312141.


\bibitem{LS} A. LeClair,and G. Sierra,
``Renormalization group limit-cycles and field theories 
for elliptic S-matrices'', Theor. Exp. (2004) P08004; hep-th/0403178. 


\bibitem{Andre} A. LeClair, 
``Interacting Bose and Fermi gases in low dimensions 
and the Riemann hypothesis'';  math-ph/0611043. 
 



\bibitem{Galindo} A. Galindo and P. Pascual,
``Quantum Mechanics I, II'', 
Springer Verlag, Berlin 1990-1991. 


 \bibitem{Neumann} J. von Neumann, 
``Allgemeine Eigenwerttheorie Hermitescher Funktionaloperatoren'',  
Math. Ann. 102, 49-131, (1929).
     

\bibitem{Twamley} J. Twamley  and G. J. Milburn, 
``The quantum Mellin transform'', New J. Phys. 8 (2006) 328. 




\bibitem{khuri} N.N. Khuri, 
``Inverse Scattering, the Coupling Constant Spectrum, and 
the Riemann Hypothesis'', Math. Phys. Anal. Geom. 5 (2002) 1-63; 
hep-th/0111067. 


\bibitem{Chadan} K. Chadan and M. Musette, 
"On an interesting singular potential", 
C.R. Acad. Sci. Paris 316 II, 1 (1993). 




\bibitem{neumann-wigner} J. von Neumann and E. P. Wigner, 
``\"Uber Merkw\"urdige Diskrete Eigenwerte'', Z. Phys. 30 (1929)
465-467. 

\bibitem{simon} B. Simon, ``On positive eigenvalues of one body
Schr\"odinger operators'', Comm. Pure. Appl. Math. 12 (1969) 531-538. 


\bibitem{arai}  M. Arai and  J. Uchiyama, 
``On the von Neumann and Wigner Potentials'', 
J. Diff Eq. 157 (1999) 348-372. 

\bibitem{cruz} J. Cruz-Sampedro, I. Herbst, R. Mart\'{\i}nez-Avenda\~no, 
``Perturbations of the Wigner-Von Neumann potential leaving 
the embedded eigenvalue fixed'', Annales Henri Poincar 3 (2002), 331-345. 


\bibitem{GR} L. S. Gradshteyn and I.M. Ryzhik, 
``Table of integrals, series and products'',  6th ed. 
Academic Press, 2000, London. 




\bibitem{Titchmarsh} E.C. Titchmarsh, ``Introduction
to the theory of Fourier integrals'', 
Oxford University Press (1937), 2nd ed.  New York. 


\bibitem{AW} G.B. Arfken and H.J. Weber, 
``Mathematical Methods for Physicist'',
Elsevier Academic Press (2005),6th ed. Oxford. 


\bibitem{enciclopedia} ``Encyclopedic Dictionary of Mathematics'', vol I,
 MIT Press (1977). The Mathematical Society of Japan. 


\bibitem{Duren}, P.L. Duren,  
``Theory of $H^p$ Spaces''. Academic Press, New York, 1970. 

\bibitem{Burnol1} J. F. Burnol, 
``An adelic causality problem related to abelian L-functions'', 
J. Number Theory 87 (2001), no. 2, 253-269; math.NT/0001013. 


\bibitem{Burnol2} J. F. Burnol,  ``On Fourier and Zeta(s)''
 Forum Mathematicum 16 (2004), 789-840; math.NT/0112254.




\bibitem{Faddeev} B.S. Pavlov and L.D. Faddeev, 
``Scattering theory and automorphic functions'', 
Sov. Math. 3, 522 (1975), Plenum Publishing Corp. translation, N.Y; 

\bibitem{Lax} Lax and R.S. Phillips, {\em Scattering Theory for Automorphic
Functions}, Princeton University Press, Princeton, 1976. 

\bibitem{G2} M.C. Gutzwiller, ``Stochastic behaviour in 
Quantum Scattering'', 
Physica {\bf D7}, 341 (1983). 


\bibitem{Joffily}  S. Joffily,  
``Jost function, prime numbers and Riemann zeta function'', 
math-ph/0303014 

\bibitem{BKL} R.K. Bhaduri, Avinash Khare, and J. Law, 
"Phase of the Riemann zeta function and the inverted harmonic oscillator", 
Physical Review E 52 no. 1 (1995) 486-491; chao-dyn/9406006.  



\bibitem{Aneva} B. Aneva, ``Symmetry of the Riemann operator'',
Phys. Lett. B 450 (1999) 388. 



\bibitem{euler-zagier-1} S. Akiyama and Y. Tanigawa, 
``Multiple zeta values at non-positive
integers'', {\em Ramanujan J.} {\bf 5} (2001), 327-351. 


\bibitem{euler-zagier-2} L. Guo and B. Zhang, 
``Renormalization of Multiple zeta values'', math.NT/0606076. 


\bibitem{Davenport} H. Davenport, ``Multiplicative Number Theory'',
2nd ed., Springer-Verlag, New York, 1980. 


\end{thebibliography}
\end{document}